\documentclass[%
reprint,
superscriptaddress,
showpacs,
amsmath,amssymb,
aps,
]{revtex4-1}

\usepackage{graphicx}% Include figure files
\usepackage{dcolumn}% Align table columns on decimal point
\usepackage{bm}% bold math
\usepackage{amsmath}
\usepackage{autobreak}
\usepackage{hyperref}% add hypertext capabilities
\usepackage[mathlines]{lineno}% Enable numbering of text and display math
\usepackage{natbib}
\usepackage{multirow}
\usepackage{graphicx}
\usepackage{float}
\usepackage{subfigure}
\usepackage{ragged2e}
\usepackage[dvipsnames]{xcolor}
\usepackage{url}
\usepackage{upgreek}

\def\GB{A^{\prime}}
\def\BmL{\rm B {-} L}
\def\nus{\nu_s}
\def\nusolar{\nu_{\odot}}
\def\KM{\rm \varepsilon}

\def\nue{\nu \mbox{-} e}
\def\nuN{\nu \mbox{-} N}

\begin{document}

\title{Search for exotic interactions of solar neutrinos in the CDEX-10 experiment}

\author{X.~P.~Geng}
\affiliation{Key Laboratory of Particle and Radiation Imaging (Ministry of Education) and Department of Engineering Physics, Tsinghua University, Beijing 100084}
\author{L.~T.~Yang}
\email{Corresponding author: yanglt@mail.tsinghua.edu.cn}
\affiliation{Key Laboratory of Particle and Radiation Imaging (Ministry of Education) and Department of Engineering Physics, Tsinghua University, Beijing 100084}
\author{Q.~Yue}
\email{Corresponding author: yueq@mail.tsinghua.edu.cn}
\affiliation{Key Laboratory of Particle and Radiation Imaging (Ministry of Education) and Department of Engineering Physics, Tsinghua University, Beijing 100084}
\author{K.~J.~Kang}
\affiliation{Key Laboratory of Particle and Radiation Imaging (Ministry of Education) and Department of Engineering Physics, Tsinghua University, Beijing 100084}
\author{Y.~J.~Li}
\affiliation{Key Laboratory of Particle and Radiation Imaging (Ministry of Education) and Department of Engineering Physics, Tsinghua University, Beijing 100084}

\author{H.~P.~An}
\affiliation{Key Laboratory of Particle and Radiation Imaging (Ministry of Education) and Department of Engineering Physics, Tsinghua University, Beijing 100084}
\affiliation{Department of Physics, Tsinghua University, Beijing 100084}

\author{Greeshma~C.}
\altaffiliation{Participating as a member of TEXONO Collaboration}
\affiliation{Institute of Physics, Academia Sinica, Taipei 11529}

\author{J.~P.~Chang}
\affiliation{NUCTECH Company, Beijing 100084}

\author{Y.~H.~Chen}
\affiliation{YaLong River Hydropower Development Company, Chengdu 610051}
\author{J.~P.~Cheng}
\affiliation{Key Laboratory of Particle and Radiation Imaging (Ministry of Education) and Department of Engineering Physics, Tsinghua University, Beijing 100084}
\affiliation{College of Nuclear Science and Technology, Beijing Normal University, Beijing 100875}
\author{W.~H.~Dai}
\affiliation{Key Laboratory of Particle and Radiation Imaging (Ministry of Education) and Department of Engineering Physics, Tsinghua University, Beijing 100084}
\author{Z.~Deng}
\affiliation{Key Laboratory of Particle and Radiation Imaging (Ministry of Education) and Department of Engineering Physics, Tsinghua University, Beijing 100084}
\author{C.~H.~Fang}
\affiliation{College of Physics, Sichuan University, Chengdu 610065}

\author{H.~Gong}
\affiliation{Key Laboratory of Particle and Radiation Imaging (Ministry of Education) and Department of Engineering Physics, Tsinghua University, Beijing 100084}
\author{Q.~J.~Guo}
\affiliation{School of Physics, Peking University, Beijing 100871}
\author{X.~Y.~Guo}
\affiliation{YaLong River Hydropower Development Company, Chengdu 610051}
\author{L.~He}
\affiliation{NUCTECH Company, Beijing 100084}
\author{S.~M.~He}
\affiliation{YaLong River Hydropower Development Company, Chengdu 610051}
\author{J.~W.~Hu}
\affiliation{Key Laboratory of Particle and Radiation Imaging (Ministry of Education) and Department of Engineering Physics, Tsinghua University, Beijing 100084}
\author{H.~X.~Huang}
\affiliation{Department of Nuclear Physics, China Institute of Atomic Energy, Beijing 102413}
\author{T.~C.~Huang}
\affiliation{Sino-French Institute of Nuclear and Technology, Sun Yat-sen University, Zhuhai 519082}
\author{H.~T.~Jia}
\affiliation{College of Physics, Sichuan University, Chengdu 610065}
\author{X.~Jiang}
\affiliation{College of Physics, Sichuan University, Chengdu 610065}

\author{S.~Karmakar}
\altaffiliation{Participating as a member of TEXONO Collaboration}
\affiliation{Institute of Physics, Academia Sinica, Taipei 11529}

\author{H.~B.~Li}
\altaffiliation{Participating as a member of TEXONO Collaboration}
\affiliation{Institute of Physics, Academia Sinica, Taipei 11529}
\author{J.~M.~Li}
\affiliation{Key Laboratory of Particle and Radiation Imaging (Ministry of Education) and Department of Engineering Physics, Tsinghua University, Beijing 100084}
\author{J.~Li}
\affiliation{Key Laboratory of Particle and Radiation Imaging (Ministry of Education) and Department of Engineering Physics, Tsinghua University, Beijing 100084}
\author{Q.~Y.~Li}
\affiliation{College of Physics, Sichuan University, Chengdu 610065}
\author{R.~M.~J.~Li}
\affiliation{College of Physics, Sichuan University, Chengdu 610065}
\author{X.~Q.~Li}
\affiliation{School of Physics, Nankai University, Tianjin 300071}
\author{Y.~L.~Li}
\affiliation{Key Laboratory of Particle and Radiation Imaging (Ministry of Education) and Department of Engineering Physics, Tsinghua University, Beijing 100084}
\author{Y.~F.~Liang}
\affiliation{Key Laboratory of Particle and Radiation Imaging (Ministry of Education) and Department of Engineering Physics, Tsinghua University, Beijing 100084}
\author{B.~Liao}
\affiliation{College of Nuclear Science and Technology, Beijing Normal University, Beijing 100875}
\author{F.~K.~Lin}
\altaffiliation{Participating as a member of TEXONO Collaboration}
\affiliation{Institute of Physics, Academia Sinica, Taipei 11529}
\author{S.~T.~Lin}
\affiliation{College of Physics, Sichuan University, Chengdu 610065}
\author{J.~X.~Liu}
\affiliation{Key Laboratory of Particle and Radiation Imaging (Ministry of Education) and Department of Engineering Physics, Tsinghua University, Beijing 100084}
\author{S.~K.~Liu}
\affiliation{College of Physics, Sichuan University, Chengdu 610065}
\author{Y.~D.~Liu}
\affiliation{College of Nuclear Science and Technology, Beijing Normal University, Beijing 100875}
\author{Y.~Liu}
\affiliation{College of Physics, Sichuan University, Chengdu 610065}
\author{Y.~Y.~Liu}
\affiliation{College of Nuclear Science and Technology, Beijing Normal University, Beijing 100875}
\author{Z.~Z.~Liu}
\affiliation{Key Laboratory of Particle and Radiation Imaging (Ministry of Education) and Department of Engineering Physics, Tsinghua University, Beijing 100084}
\author{H.~Ma}
\affiliation{Key Laboratory of Particle and Radiation Imaging (Ministry of Education) and Department of Engineering Physics, Tsinghua University, Beijing 100084}
\author{Y.~C.~Mao}
\affiliation{School of Physics, Peking University, Beijing 100871}
\author{Q.~Y.~Nie}
\affiliation{Key Laboratory of Particle and Radiation Imaging (Ministry of Education) and Department of Engineering Physics, Tsinghua University, Beijing 100084}
\author{J.~H.~Ning}
\affiliation{YaLong River Hydropower Development Company, Chengdu 610051}
\author{H.~Pan}
\affiliation{NUCTECH Company, Beijing 100084}
\author{N.~C.~Qi}
\affiliation{YaLong River Hydropower Development Company, Chengdu 610051}
\author{J.~Ren}
\affiliation{Department of Nuclear Physics, China Institute of Atomic Energy, Beijing 102413}
\author{X.~C.~Ruan}
\affiliation{Department of Nuclear Physics, China Institute of Atomic Energy, Beijing 102413}

\author{Z.~She}
\affiliation{Key Laboratory of Particle and Radiation Imaging (Ministry of Education) and Department of Engineering Physics, Tsinghua University, Beijing 100084}
\author{M.~K.~Singh}
\altaffiliation{Participating as a member of TEXONO Collaboration}
\affiliation{Institute of Physics, Academia Sinica, Taipei 11529}
\affiliation{Department of Physics, Banaras Hindu University, Varanasi 221005}
\author{T.~X.~Sun}
\affiliation{College of Nuclear Science and Technology, Beijing Normal University, Beijing 100875}
\author{C.~J.~Tang}
\affiliation{College of Physics, Sichuan University, Chengdu 610065}
\author{W.~Y.~Tang}
\affiliation{Key Laboratory of Particle and Radiation Imaging (Ministry of Education) and Department of Engineering Physics, Tsinghua University, Beijing 100084}
\author{Y.~Tian}
\affiliation{Key Laboratory of Particle and Radiation Imaging (Ministry of Education) and Department of Engineering Physics, Tsinghua University, Beijing 100084}
\author{G.~F.~Wang}
\affiliation{College of Nuclear Science and Technology, Beijing Normal University, Beijing 100875}
\author{L.~Wang}
\affiliation{Department of  Physics, Beijing Normal University, Beijing 100875}
\author{Q.~Wang}
\affiliation{Key Laboratory of Particle and Radiation Imaging (Ministry of Education) and Department of Engineering Physics, Tsinghua University, Beijing 100084}
\affiliation{Department of Physics, Tsinghua University, Beijing 100084}
\author{Y.~F.~Wang}
\affiliation{Key Laboratory of Particle and Radiation Imaging (Ministry of Education) and Department of Engineering Physics, Tsinghua University, Beijing 100084}
\author{Y.~X.~Wang}
\affiliation{School of Physics, Peking University, Beijing 100871}
\author{H.~T.~Wong}
\altaffiliation{Participating as a member of TEXONO Collaboration}
\affiliation{Institute of Physics, Academia Sinica, Taipei 11529}
\author{S.~Y.~Wu}
\affiliation{YaLong River Hydropower Development Company, Chengdu 610051}
\author{Y.~C.~Wu}
\affiliation{Key Laboratory of Particle and Radiation Imaging (Ministry of Education) and Department of Engineering Physics, Tsinghua University, Beijing 100084}
\author{H.~Y.~Xing}
\affiliation{College of Physics, Sichuan University, Chengdu 610065}
\author{R. Xu}
\affiliation{Key Laboratory of Particle and Radiation Imaging (Ministry of Education) and Department of Engineering Physics, Tsinghua University, Beijing 100084}
\author{Y.~Xu}
\affiliation{School of Physics, Nankai University, Tianjin 300071}
\author{T.~Xue}
\affiliation{Key Laboratory of Particle and Radiation Imaging (Ministry of Education) and Department of Engineering Physics, Tsinghua University, Beijing 100084}
\author{Y.~L.~Yan}
\affiliation{College of Physics, Sichuan University, Chengdu 610065}

\author{N.~Yi}
\affiliation{Key Laboratory of Particle and Radiation Imaging (Ministry of Education) and Department of Engineering Physics, Tsinghua University, Beijing 100084}
\author{C.~X.~Yu}
\affiliation{School of Physics, Nankai University, Tianjin 300071}
\author{H.~J.~Yu}
\affiliation{NUCTECH Company, Beijing 100084}
\author{J.~F.~Yue}
\affiliation{YaLong River Hydropower Development Company, Chengdu 610051}
\author{M.~Zeng}
\affiliation{Key Laboratory of Particle and Radiation Imaging (Ministry of Education) and Department of Engineering Physics, Tsinghua University, Beijing 100084}
\author{Z.~Zeng}
\affiliation{Key Laboratory of Particle and Radiation Imaging (Ministry of Education) and Department of Engineering Physics, Tsinghua University, Beijing 100084}
\author{B.~T.~Zhang}
\affiliation{Key Laboratory of Particle and Radiation Imaging (Ministry of Education) and Department of Engineering Physics, Tsinghua University, Beijing 100084}
\author{F.~S.~Zhang}
\affiliation{College of Nuclear Science and Technology, Beijing Normal University, Beijing 100875}
\author{L.~Zhang}
\affiliation{College of Physics, Sichuan University, Chengdu 610065}
\author{Z.~H.~Zhang}
\affiliation{Key Laboratory of Particle and Radiation Imaging (Ministry of Education) and Department of Engineering Physics, Tsinghua University, Beijing 100084}
\author{Z.~Y.~Zhang}
\affiliation{Key Laboratory of Particle and Radiation Imaging (Ministry of Education) and Department of Engineering Physics, Tsinghua University, Beijing 100084}
\author{K.~K.~Zhao}
\affiliation{College of Physics, Sichuan University, Chengdu 610065}
\author{M.~G.~Zhao}
\affiliation{School of Physics, Nankai University, Tianjin 300071}
\author{J.~F.~Zhou}
\affiliation{YaLong River Hydropower Development Company, Chengdu 610051}
\author{Z.~Y.~Zhou}
\affiliation{Department of Nuclear Physics, China Institute of Atomic Energy, Beijing 102413}
\author{J.~J.~Zhu}
\affiliation{College of Physics, Sichuan University, Chengdu 610065}

\collaboration{CDEX Collaboration}
\noaffiliation

\author{Y.~C.~Wu}
\affiliation{Department of Physics and Institute of Theoretical Physics, Nanjing Normal University, Nanjing, 210023}

\date{\today}

\begin{abstract}
We investigate exotic neutrino interactions using the 205.4 kg$\cdot$day dataset from the CDEX-10 experiment at the China Jinping Underground Laboratory. New constraints on the mass and couplings of new gauge bosons are presented. Two nonstandard neutrino interactions are considered: a $U(1)_{\BmL}$ gauge-boson-induced interaction between an active neutrino and electron/nucleus, and a dark-photon-induced interaction between a sterile neutrino and electron/nucleus via kinetic mixing with a photon. This work probes an unexplored parameter space involving sterile neutrino coupling with a dark photon. New laboratory limits are derived on dark photon masses below $1~{\rm eV}/c^{2}$ at some benchmark values of $\Delta m_{41}^{2}$ and $g^{\prime2}{\rm{sin}}^{2}2\theta_{14}$. 

\end{abstract}

\maketitle

%\emph{Introduction.}—
\section{\label{sec1}Introduction}
Various cosmological and astrophysical observations at different scales reveal phenomena beyond the Standard Model (SM)~\cite{cosmology}. The measurement of nonstandard interaction (NSI) in the neutrino sector is an attractive approach to probe beyond-SM physics~\cite{Proceedings:2019qno,Ohlsson_2013}. Current experimental efforts on neutrino NSI are conducted with different neutrino sources, such as reactor neutrinos~\cite{NEOS:PRL2017,Neutrino4:PRD2021,STEREO:PRD2020,PROSPECT:PRL2018,DANSS:PLB2018,Solid:JINST2021,DayaBay:PRL2014,RENO:PRL2020,JUNO:2020ijm}, accelerator neutrinos~\cite{MiniBooNE:PRD2021,LSND:PRD2001,MicroBooNE,JSNS2:NIMA2021}, and radioactive sources~\cite{GALLEX:PLB1998,SAGE:PRC1999,BEST:PRL2022,BEST1,BEST2}. 
In addition to these terrestrial sources, NSI can also be probed with neutrinos from astrophysical sources, such as stars~\cite{solarneutrino}, supernovae~\cite{SupernovaNeutrino,SupernovaNeutrino2}, terrestrial atmosphere~\cite{Kajita:2012vc}, and others~\cite{RevModPhys84}.
In this paper, we investigate two attractive exotic neutrino NSIs, where new gauge boson mediators (generically denoted as $\GB$) from the hidden sector couple active or sterile neutrinos with SM particles. Constraints are placed with data from the CDEX-10 experiment~\cite{cdex102018,cdex_darkphoton,cdex10_eft,CRDM,CDEX_DM_e,cdex10_tech} using solar neutrino ($\nusolar$) as the source.

The first NSI model is based on a gauged $U(1)_{\BmL}$ symmetry~\cite{newphysics,U1BL} with the corresponding $\GB$ interacting with SM particles with a nonzero $\BmL$ number (baryon number minus lepton number) at tree level. This global $U(1)$ symmetry appears in grand unified theory and will not be violated by chiral and gravitational anomalies. The symmetry can give rise to neutrino mass when spontaneously broken, and the corresponding $\GB$ is a dark matter (DM) candidate.
The free parameters are the new gauge coupling constant ($g_{\BmL}$) and the gauge boson mass ($M_{\GB}$). This additional $U(1)_{\BmL}$ mediator would lead to a new interaction between the neutrino and the SM particles which is measurable by enhanced event rates. 
The second NSI model considers the existence of a sterile neutrino ($\nus$) which couples with $\GB$, called a dark photon, under a new gauged symmetry $U(1)^\prime$~\cite{newphysics}.
The dark photon is a popular DM candidate and can be a portal between the SM and the dark sector. Observable interactions between $\nus$ and SM matter are induced by $\GB$. The coupling strength between $\GB$ and $\nus$ is parametrized by $g^\prime$, while that with SM particles with charge $Q$ is via its kinetic mixing ($\KM$) with the SM photons.
Other interesting NSI models with an extra $U(1)$ gauge boson~\cite{Fayet:1980ad,Fayet:1990wx} are beyond the scope of this work.

The $p$-type point contact germanium (PPCGe) semiconductor in ionization mode is ideal for the studies of exotic processes due to its ultralow energy threshold of $\mathcal{O}$(100 eVee) (``eVee'' represents the electron equivalent energy derived from energy calibration) and low background level of $\mathcal{O}$(1 count kg$^{-1}$ keVee$^{-1}$ day$^{-1}$)~\cite{Barbeau_2007,soma2016}. It has been adopted by the CDEX experiment~\cite{cdex0,cdex1,cdex12014,cdex12016,cdex1b2018,cdex1b_am,cdex102018,cdex10_tech,cdex10_eft,cdex_darkphoton,CRDM,CDEX_DM_e} for searches of DM and beyond-SM NSI at the China Jinping Underground Laboratory (CJPL) where the rock overburden is about 2400 m~\cite{cjpl}. The second phase of the CDEX experiment, CDEX-10, takes data with a 10-kg PPCGe detector array, consisting of three triple-element PPCGe detector strings encapsulated in copper vacuum tubes and immersed in liquid nitrogen which serves both for cooling and shielding.
The CDEX-10 experimental configuration was described in Refs.~\cite{cdex102018,cdex10_tech}. Data taking started in February 2017, and the physics analysis threshold is 160 eVee~\cite{cdex102018}. Previous scientific results were published in Refs.~\cite{cdex102018,cdex_darkphoton,cdex10_eft,CRDM,CDEX_DM_e}.

%\emph{Data analysis.}—
\section{\label{sec2}Data analysis}
\begin{figure}[!tbp]
	\centering
	\includegraphics[width=\columnwidth]{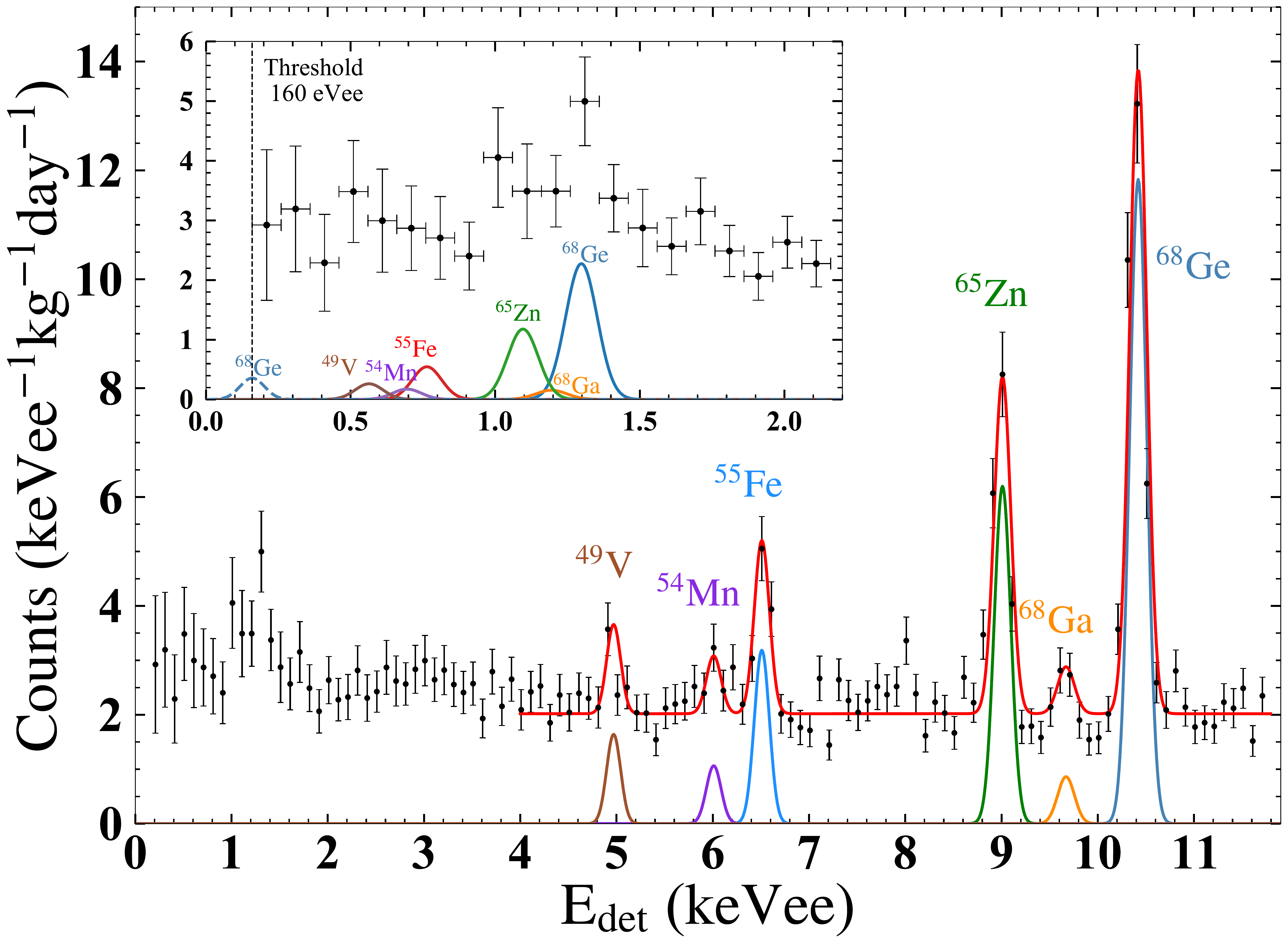}
	\caption{The measured energy spectrum with error bars including both the statistical and systematical uncertainties based on the 205.4 kg$\cdot$day dataset of the CDEX-10 experiment~\cite{cdex_darkphoton,cdex10_eft,CRDM,CDEX_DM_e}. The bin width is 100 eVee and the energy range is 0.16--11.76 keVee. The characteristic K-shell x-ray peaks from internal cosmogenic radionuclides are marked by the isotope symbols in color. Both the best fit curve of the measured energy spectrum in 4--11.8 keV, which is the red line, and the contributions of these radionuclides derived by the best fit are superimposed. Displayed in the inset are the contributions of L- and M-shell x-ray peaks derived from the corresponding K-shell line intensities~\cite{K-X-ray}. The L-shell x-ray peaks are shown in solid lines. The dashed line represents the M-shell x-ray peak of $^{68}$Ge.}
	\label{fig:spectrum}
\end{figure}

\begin{figure}[!tbp]
	\includegraphics[width=\columnwidth]{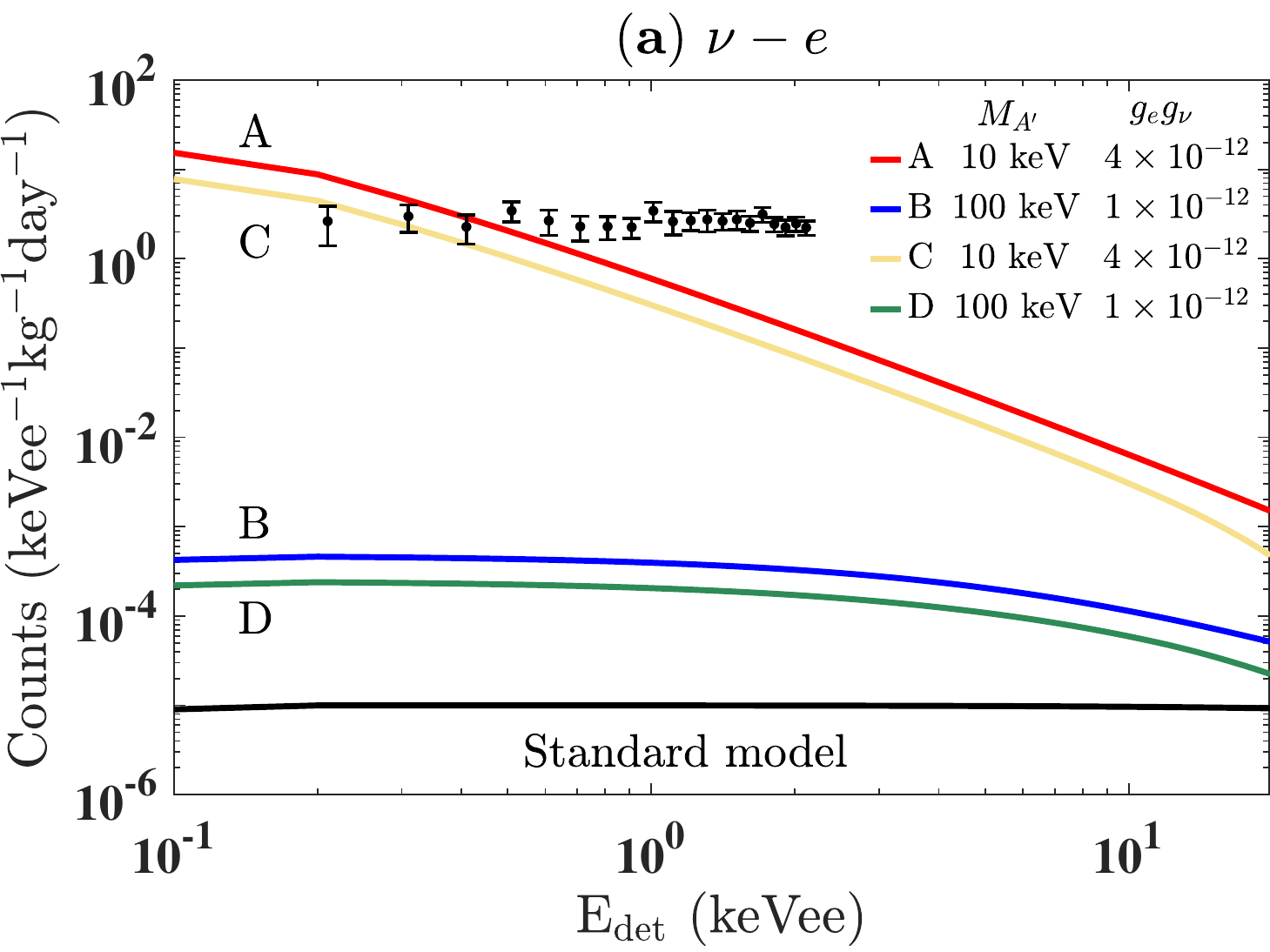}
	\includegraphics[width=\columnwidth]{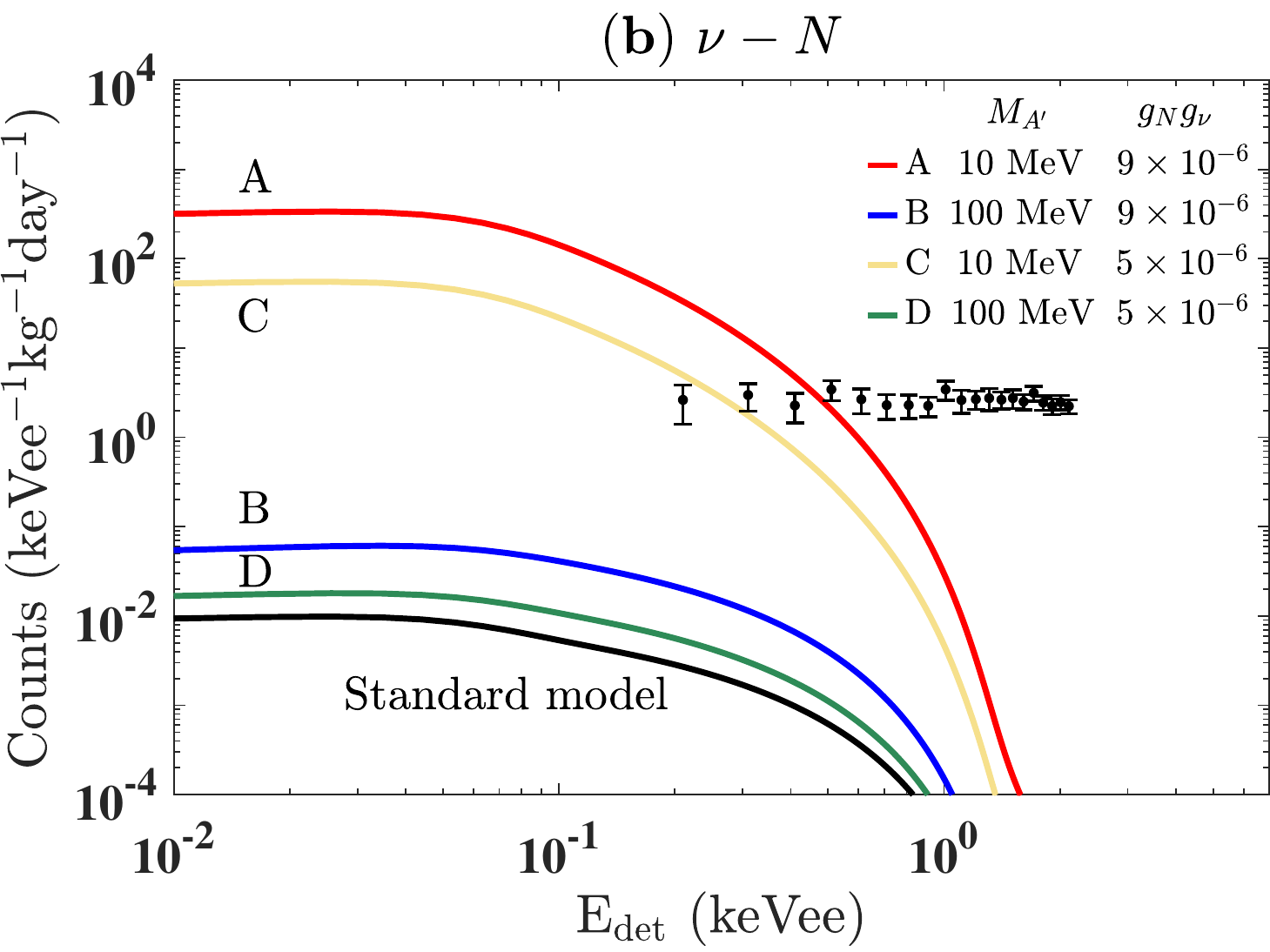}
	\caption{The measured (black data points with uncertainties) and expected (colored lines) event rates under the two neutrino NSI models discussed in the text for (a) $\nue$ and (b) $\nuN$ scatterings. Lines A and B correspond to benchmark parameter choices in model I, while C and D are for model II. For lines C and D, the parameter $\sin2\theta_{14}$ has been absorbed into the $g_{e}g_{\nu}$ and $g_{N}g_{\nu}$. The energy resolution of CDEX-10~\cite{cdex102018,cdex10_tech,CRDM,CDEX_DM_e} was considered in the evaluation of the expected event rates. The quenching factor in Ge for $\nuN$ is calculated with the $\tt TRIM$ package~\cite{TRIM}. The CDEX-10 data corresponds to the residual spectra with the L- and M-shell x-ray contributions subtracted in 0.16--2.16 keVee, at a bin width of 100 eVee~\cite{cdex102018,CRDM,CDEX_DM_e}.}
	\label{fig:gBL}
\end{figure}

The data analysis of this work is based on a 205.4 kg$\cdot$day dataset from CDEX-10~\cite{cdex_darkphoton,cdex10_eft,CRDM,CDEX_DM_e} and follows the established procedures of previous works~\cite{cdex12016,cdex102018,cdex10_tech,cdex_darkphoton,cdex1b2018}. The energy calibration was performed with zero energy (defined by random trigger events) and the internal cosmogenic K-shell x-ray peaks: 8.98 keVee of $\rm ^{65}Zn$ and 10.37 keVee of $\rm ^{68,71}Ge$. The signal events are identified after pedestal noise cut, physics events selection, and bulk/surface events discrimination~\cite{Li:2014a,Yang:2018a}. The measured energy spectrum in the detector ($E_{det}$) in keVee units after physics event selections and efficiency corrections is shown in Fig.~\ref{fig:spectrum}. The physics analysis threshold is set to be 160 eVee at which the combined signal efficiency (including the trigger efficiency and the efficiency for the pulse shape discrimination) is 4.5\%~\cite{cdex10_tech}. The characteristic K-shell x-ray peaks from internal cosmogenic radionuclides like $\rm^{68}Ge$, $\rm^{68}Ga$, $\rm^{65}Zn$, $\rm^{55}Fe$, $\rm^{54}Mn$, and $\rm^{49}V$ can be identified. Their intensities are derived from the best fit of the spectrum~\cite{cdex102018}. At the sub-keVee energy range relevant to this analysis, background events are dominated by Compton scattering of high-energy gamma rays and internal radioactivity from long-lived cosmogenic isotopes. Figure~\ref{fig:gBL} shows the residual spectrum in the region of 0.16--2.16 keVee after subtracting the contributions from L- and M-shell x-ray peaks which are derived from the corresponding K-shell line intensities~\cite{cdex102018,cdex_darkphoton,cdex10_eft,CRDM,CDEX_DM_e}. This is illustrated in the inset of Fig.~\ref{fig:spectrum}. The count rate is several orders of magnitude larger than the predictions of SM $\nu_\odot$ interaction.

A minimum-$\chi^{2}$ analysis~\cite{cdex102018,cdex_darkphoton,cdex12014,CDEX_DM_e} is applied to the residual spectrum in the range 0.16--2.16 keVee, in which $\chi^{2}$ is defined as
\begin{align}
	\label{equ:chi-square}
	\chi^{2} = \sum_{i}\frac{[n_{i} - S_{i} - B]^{2}}{\Delta_{i}^{2}},
\end{align}
where $n_{i}$ is the measured count at the $i$th energy bin, and $S_{i}$ is the expected event rate due to the neutrino NSI model being probed. $\Delta_{i}$ is the combination of the statistical and systematic uncertainties~\cite{cdex102018}, and $B$ is the flat background contribution from the Compton scattering of high-energy gamma rays. The best estimator of the couplings (see discussion below) at certain mediator mass $M_{A'}$ is evaluated by minimizing the $\chi^{2}$ values. Upper limits at 90\% confidence level (CL) are derived by the unified approach~\cite{fcmethod}.  

%\emph{Framework of exotic neutrino interactions.}—
\section{\label{sec3}Framework of exotic neutrino interactions}
Within the SM, neutrinos can interact with Ge and produce detectable electronic and nuclear recoils via elastic $\nue$ and $\nuN$ electroweak interactions, respectively. A popular model for exotic neutrino interactions introduces a new gauge boson mediator described by
\begin{align}
	\label{equ:interactions}
	\mathcal{L}_{\rm int}\supset g_e \bar e \gamma^\mu e A'_\mu + g_q \bar q\gamma^\mu q A'_\mu +g_\nu \bar\nu\gamma^\mu P_{L,R}\nu A'_\mu
\end{align}
where $\GB$ is the extra mediator with mass $M_{\GB}$ from a $U(1)$ gauge group, $\nu$ can be either an active or sterile neutrino, and $g_{e,q,\nu}$ are the couplings between the $\GB$ with the corresponding fermions.
The neutrino-induced scattering rate is
\begin{equation} \label{con:nmmE}
	\begin{aligned}
		\frac{dR}{dE_{r}} = N_{T}\times\int_{E_{\nu}^{min}}^{\infty}\frac{d\Phi}{dE_{\nu}}\frac{d\sigma}{dE_{r}}dE_{\nu},
	\end{aligned}
\end{equation}
where $N_{T}$ is the number of target nuclei, or electrons, per unit of mass of the detector material (for $\nuN$ and $\nue$, respectively), and $E_{\nu}^{min}$ is the minimum neutrino energy required to generate recoil energy $E_r$. $\frac{d\sigma}{dE_{r}}$ is the differential cross section, and $\frac{d\Phi}{dE_{\nu}}$ is the differential flux of neutrinos. The B16-GS98 solar model (also referred to as the high-metallicity, or HZ model) is adopted and the values for the $\nusolar$ fluxes are taken from Ref.~\cite{Vinyoles2016}. Following Eq. (\ref{equ:interactions}), the enhancements of the $\nue$ and $\nuN$ scattering cross section are given by, respectively,
\begin{align}
	\label{equ:e-nu-dsigma}
	\frac{d\sigma(\nu e\to \nu e)}{dE_r} &=  \frac{(g_e g_\nu)^2m_e}{4\pi p_\nu^2(M_{A'}^2+2E_rm_e)^2}\nonumber\\
	\times&\left[2E_\nu^2+E_r^2-2E_rE_\nu -E_rm_e-m_\nu^2\right],\\
	\label{equ:N-nu-dsigma}
	\frac{d\sigma(\nu N\to \nu N)}{dE_r} &= \frac{ (g_N g_\nu)^2
		m_NF^2(E_r)}{4\pi p_\nu^2(M_{A'}^2+2E_rm_N)^2}\nonumber\\
	\times&\left[2E_\nu^2+E_r^2-2E_rE_\nu-E_rm_N-m_\nu^2\right],
\end{align}
where $E_r$ is the recoil energy of the target, $E_\nu$ is the neutrino energy, $m_{e,N,\nu}$ are the masses of the electron, target nucleus, and neutrino, $M_{\GB}$ is the mass of the extra gauge boson, $g_N$ is the coherent coupling of the gauge boson with the nucleus $g_N = g_p Z + g_n(A-Z)$ with $g_p = 2g_u + g_d$ and $g_n = g_u + 2g_d$, and $F^2(E_r)$ is the nuclear form factor which describes the loss of coherence due to the internal structure of the nucleus. The conventional Helm form factor~\cite{form,formfactor} is adopted in this analysis. The observed energy deposition $E_{det}$ 
in $\nue$ is equal to the actual electron recoil energy $E_r$. 
For $\nuN$, the observed total deposited energy $E_{det}$ is different from the actual nuclear recoil energy $E_r$ and should be corrected by $E_{det}=Q_{nr}E_r$, where the quenching factor $Q_{nr}$ in Ge is calculated by the $\tt TRIM$ package~\cite{TRIM}.
The differential event rates in Ge for the $\nue$ and $\nuN$ scattering from $\nusolar$ at several benchmark physics parameters are evaluated and displayed in Figs.~\ref{fig:gBL}(a) and ~\ref{fig:gBL}(b), respectively.

Within this framework, two neutrino NSI models are studied and limits are derived with the CDEX-10 data. 

%\emph{Model I: Active neutrinos and SM particles coupled through $U(1)_{\BmL}$ gauge boson.}— 
\subsection{\label{sec2a}Model I: Active neutrinos and SM particles coupled through the $U(1)_{\BmL}$ gauge boson}
The coupling of the $\GB$ to SM particles gives rise to additional contributions to the $\nue$ and $\nuN$ differential cross sections. The additional contribution can be classified into two categories: pure contribution from the $U(1)_{\BmL}$ gauge boson and the interference term between the $U(1)_{\BmL}$ gauge boson and the SM~\cite{U1BL}. Contributions of the interference terms should be considered when the effects due to new physics are small or comparable relative to the SM contribution. This is the case applicable to experiments where the SM cross sections are measured, such as TEXONO-CsI~\cite{TEXONO_CsI}, Borexino~\cite{Borexino_exclusion}, XMASS~\cite{XMASS}, and CHARM II~\cite{VILAIN1993,VILAIN1994}. Otherwise, when the ranges of new physics effects are large compared to the SM, the interference term can in general be neglected~\cite{U1BL}, and as the event rates measured in the CDEX-10 experiment~\cite{cdex102018,cdex_darkphoton,cdex10_eft,CRDM,CDEX_DM_e} are much larger than the SM prediction, it is safe to ignore the interference contribution in this work. For the pure contribution, the induced NSI is universal and couples to the $\BmL$ number of each particle: $3g_q = -g_e = -g_\nu = g_{\BmL}$. The active $\nusolar$'s are considered, and the approximation $m_\nu \approx 0$ is made. 

\begin{figure}[!tbp]
	\centering
	\includegraphics[width=\columnwidth]{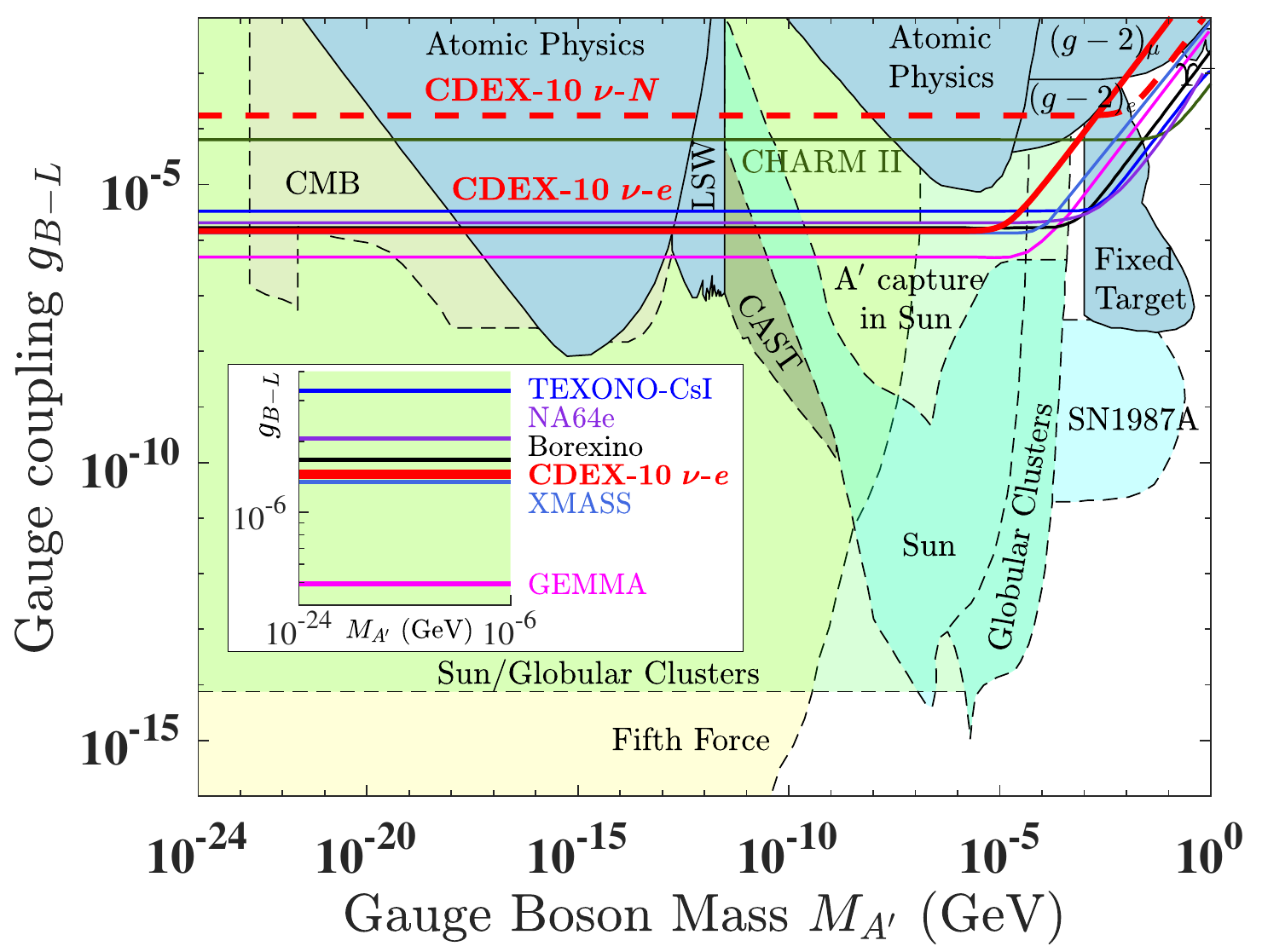}
	\caption{Constraints on a $U(1)_{\BmL}$ gauge boson with coupling $g_{\BmL}$ and mass $M_{\GB}$. The 90\% CL limits from CDEX-10 $\nusolar$ analysis are shown in red, where the solid and dashed lines represent $\nue$ and $\nuN$ scattering, respectively. Other laboratory constraints from $\nue$ scattering are displayed, with the inset in expanded scale, showing limits from reactor neutrinos at TEXONO~\cite{TEXONO_CsI} and GEMMA~\cite{GEMMA}, solar neutrinos at Borexino~\cite{Borexino_exclusion} and XMASS~\cite{XMASS}, and accelerator neutrino beams at CHARM II~\cite{VILAIN1993,VILAIN1994} and the fixed-target experiment NA64~\cite{NA64e}. Excluded regions from astrophysics analysis~\cite{Sun1,Sun2,CAST,CMB1,CMB2,SN1987A,ENERGY_LOSS}, typically model dependent, are depicted in light shade. The other constraints are from Refs.~\cite{U1BL,LSW,atomic_physics,fifth_force,fixed_1,fixed_2,fixed_3,g-21,g-22,Upsilon}.
}
	\label{fig:gblS}
\end{figure}

The expected event rates for $\nusolar \mbox{-} e$ and 
$\nusolar \mbox{-} N$ scattering with benchmark parameters (cases A and B) are displayed in Figs.~\ref{fig:gBL}(a) and \ref{fig:gBL}(b), respectively, and compared with the measured CDEX-10 data. Cross-sections enhancement between the two cases follows the $1/(M_{A^{\prime}}^{2}+2E_{r}m_{e/N})^{2}$ dependence in Eqs. (\ref{equ:e-nu-dsigma}) and (\ref{equ:N-nu-dsigma}). At light $\GB$ region where $M_{\GB} \ll \sqrt{E_r m_{e/N}}$, the energy spectra scale as $\propto E_{r}^{-2}$~\cite{newphysics}, contributing to enhanced event rates for low threshold experiments.
Upper limits from both channels at 90\% CL on $g_{\BmL}$ as a function of $M_{\GB}$ derived with the unified approach~\cite{fcmethod} are depicted in Fig.~\ref{fig:gblS}, with constraints from previous laboratory experiments~\cite{U1BL,GEMMA,Borexino_exclusion,TEXONO_CsI,VILAIN1993,VILAIN1994,NA64e,XMASS} and model-dependent astrophysical bounds~\cite{Sun1,Sun2,CAST,CMB1,CMB2,SN1987A,ENERGY_LOSS} superimposed. The $\nue$ scattering gives better sensitivity at $M_{\GB} < 2~{\rm MeV}$, while $\nuN$ scattering dominates at large $M_{\GB}$. This is due to kinematics constraints as well as $ m_e \ll m_N $ in Eqs. (\ref{equ:e-nu-dsigma}) and (\ref{equ:N-nu-dsigma}).
This work on the CDEX-10 $\nusolar$ analysis provides limits on $U(1)_{\BmL}$ of model I: $g_{\BmL} < 1.45\times 10^{-6}$ at $M_{\GB} =  1 ~ {\rm keV}$ by $\nue$ scattering and 
$g_{\BmL} < 8.74\times 10^{-4}$ at $M_{\GB} = 10 ~ {\rm MeV}$ by $\nuN$ scattering. The $\nue$ scattering analysis places the strongest limit for the $U(1)_{\BmL}$ gauge boson with mass $M_{\GB} < 1~{\rm keV}$ among solid-state detector-based experiments that use solar neutrino as the source, which is competitive with those of current leading laboratory constraints~\cite{U1BL,TEXONO_CsI,GEMMA,Borexino_exclusion,XMASS,VILAIN1993,VILAIN1994,NA64e}.

%\emph{Model II: Sterile neutrinos and SM particles coupled through a dark photon.}—
\subsection{\label{sec2b}Model II: Sterile neutrinos and SM particles coupled through a dark photon}
The dark photon couples $\nus$ to SM particles with an extra gauged $U(1)'$ symmetry via the mixing with the photons. The $\nus$'s are singlet under the SM gauge group but charged under the $U(1)'$ symmetry. Following Eq.~(\ref{equ:interactions}), we have $g_\nu = g'$ and $g_f = \KM eQ$ for $f = e, q$, where $Q$ is the charge of the corresponding fermion, and $g^{\prime}$ is the $U(1)^{\prime}$ gauge coupling constant.

The expected event rate depends on the $\nus$ flux. In this analysis, light $\nus$ with a mass less than $\mathcal{O}(100)~{\rm keV}$ is considered. A small admixture of $\nus$ to the $\nusolar$ flux 
can be produced by oscillation on its way to Earth~\cite{newphysics}. The vacuum oscillation probability in a two-flavor approximation is given by the usual expression~\cite{newphysics}
\begin{equation} \label{con:os}
	\begin{aligned}
		P(\nu_{a}\rightarrow\nu_{s}) = {\rm{sin}}^{2}2\theta_{14}{\rm{sin}}^{2}(\frac{\Delta m_{41}^{2}L}{4E}),
	\end{aligned}
\end{equation}
where $\theta_{14}$ is the effective active-sterile neutrino mixing angle in vacuum, $\Delta m_{41}^{2} = m_{4}^{2} - m_{1}^2$ is the splitting between the squared mass of $\nus$ mass eigenstate ($m_{4}$) and the dominant $\nusolar$ mass eigenstate ($m_{1}$) in vacuum, $L$ is the distance traveled by the neutrino, and $E$ is the neutrino energy.

The expected event rate under this NSI model follows a similar pattern to that of model I as illustrated in Figs. (\ref{fig:gBL})a and (\ref{fig:gBL})b (cases C and D). The cross-section enhancement at low recoil energy ($\propto E_r^{-2}$) discussed above also applies.
We note that the extra dependence on $\sin2\theta_{14}$ is absorbed into $g_{e,N}g_\nu$. The constraints on the $\KM$ parameter depend on $\Delta m_{41}^{2}$ and $g^{\prime2}{\rm{sin}}^{2}2\theta_{14}$. Studies of these interactions with $\nusolar$ would yield much higher sensitivities to the couplings thanks to the large $L$ values compared to terrestrial sources.

\begin{figure}[!tbp]
	\centering\includegraphics[width=\columnwidth]{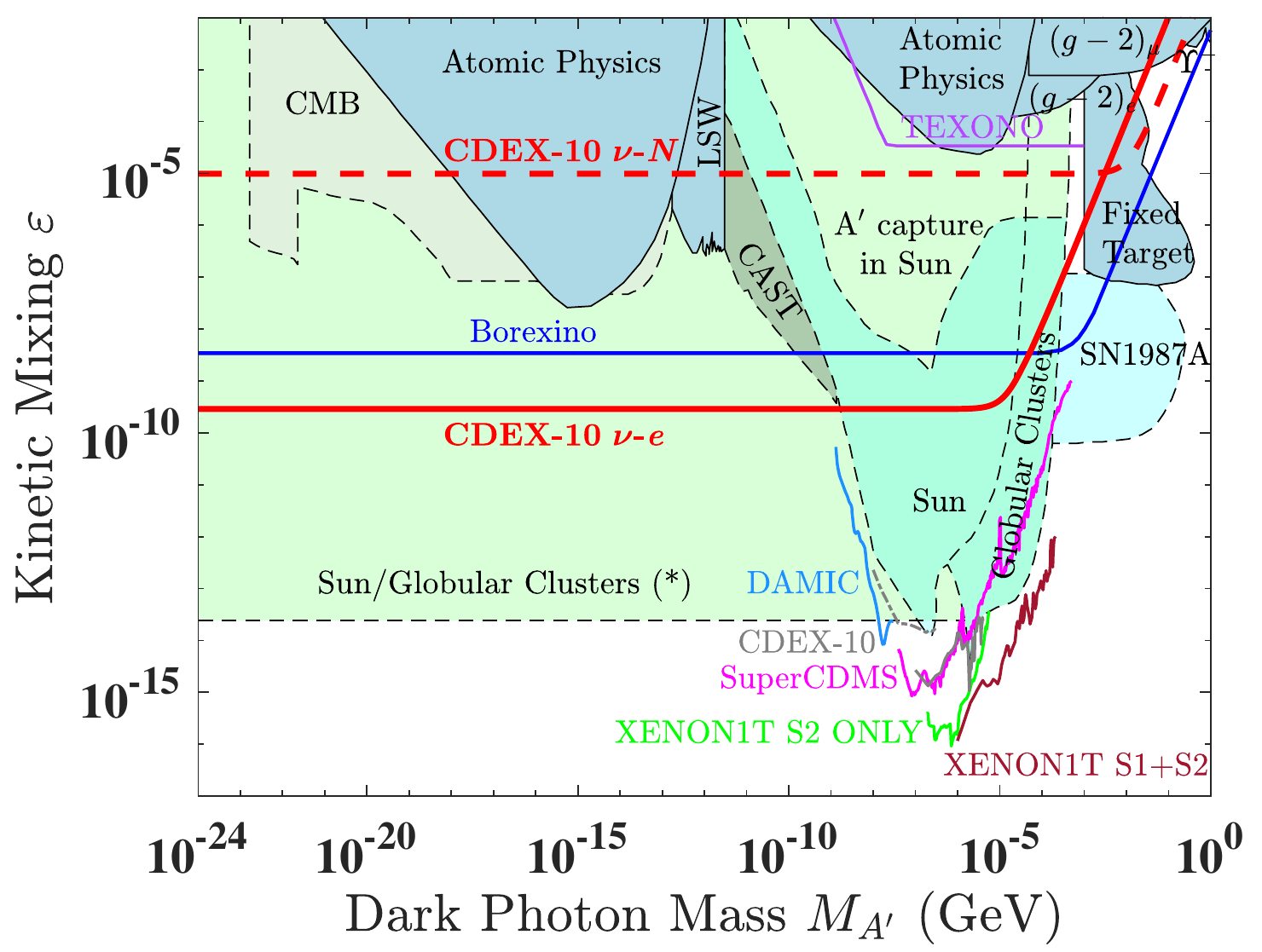}
	\caption{Constraints on light $\GB$ gauge bosons kinetically mixed with the photon as a function of the $\GB$ mass and the kinetic mixing parameter $\KM$, at the benchmark values of $\Delta m_{41}^{2} = (10~\rm keV)^{2}$ and $g^{\prime2}{\rm{sin}}^{2}2\theta_{14} = 10^{-4}$, which follow earlier phenomenological interpretations of Borexino data~\cite{Borexino_exclusion} by Ref.~\cite{newphysics}. The 90\% CL limits from CDEX-10 $\nusolar$ analysis are shown in red, where the solid and dashed lines represent $\nue$ and $\nuN$, respectively. The gray lines (CDEX-10) represent previous CDEX-10 constraints on $\epsilon$ using the same dataset under a different theoretical framework~\cite{cdex_darkphoton}: The solid and dashed lines stand for DM and solar dark photon, respectively. The Sun/Globular Clusters bounds marked (*) are valid only for $\nus \leq 10 ~ {\rm keV}$~\cite{newphysics}. Excluded regions from astrophysics analysis~\cite{newphysics,Sun1,Sun2,CAST,CMB1,CMB2,SN1987A,ENERGY_LOSS}, typically model dependent, are depicted in light shade. The other laboratory constraints are from Refs.~\cite{Borexino_exclusion,TEXONO_DP,DAMIC_DP,cdex_darkphoton,SuperCDMS_DP,XENON1T_S1_S2_DP,XENON1T_S2_DP}.
}
	\label{fig:snS}
\end{figure}

Using a minimum-$\chi^2$ analysis as discussed, no significant signal of $\nue$ or $\nuN$ scattering is observed. The 90\% CL bounds from CDEX-10 $\nusolar$ analysis are shown in Fig.~\ref{fig:snS}, at the selected parameters $\Delta m_{41}^{2} = (10~\rm keV)^{2}$ and $g^{\prime2}{\rm{sin}}^{2}2\theta_{14} = 10^{-4}$ following earlier analyses of Borexino~\cite{newphysics,Borexino_exclusion}.

Both the $\nue$ and $\nuN$ channels provide improved sensitivities to some regions of the parameter space. For $M_{\GB} < 3~\rm MeV$, the $\nue$ scattering leads to better constraints in the couplings, while for large $M_{\GB}$, the more stringent limits come from $\nuN$ scattering. The CDEX-10 results from the $\nue$ scattering improve over the Borexino bounds~\cite{newphysics,Borexino_exclusion} on the kinetic mixing parameter $\KM$ in $M_{\GB} < 50~{\rm keV}$. The upper limits at 90\% CL of $\KM < 2.89 {\times} 10^{-10}$ for $M_{\GB} = 1 \,\rm keV$ are derived for model II. The limits represent the most sensitive laboratory constraints for light dark photons with mass $M_{A^{\prime}}<1~\rm eV$. Unlike earlier experiments following conventional dark photon analysis~\cite{DAMIC_DP,SuperCDMS_DP,XENON1T_S2_DP,XENON1T_S1_S2_DP,cdex_darkphoton} where the sensitivities are limited by the detector threshold, the results from CDEX-10 under the extended model open a new window for the research of an extremely low-mass dark photon not covered by other laboratory, cosmological, and astrophysical bounds. In particular, the results from our analysis are complementary to the astrophysical bounds from the Sun/Globular Clusters which are model specific and can be evaded~\cite{newphysics} if the $\nus$'s are more massive than $\sim$10~keV or are not produced in a significant amount in the early Universe, or are chameleonlike, while our results will be robust against these variations.

%\emph{Summary.}—
\section{\label{sec4}Summary}
In this paper, we report results on the searches of nonstandard neutrino (both active or sterile) interactions with the dataset corresponding to 205.4 kg$\cdot$day exposure from the CDEX-10 experiment. No significant signal is observed, and the measured event rates are translated into upper limits on the couplings of two beyond-SM NSIs using $\nusolar$ as a probe. One model postulates a $U(1)_{\BmL}$ gauge-boson-induced interaction between active neutrinos and electron/nucleus and another a kinetically mixed dark-photon-induced interaction between sterile neutrino and electron/nucleus. The most stringent constraint among solid-state detector-based experiments that use solar neutrino as a source is placed on the $U(1)_{\BmL}$ gauge boson with mass $M_{\GB} < 1~{\rm keV}$. A new parameter space of $\KM$ on a dark photon for $M_{\GB} < 1~{\rm eV}$ at some benchmark values of $\Delta m_{41}^{2}$ and $g^{\prime2}{\rm{sin}}^{2}2\theta_{14}$ is probed. Our results extend the reach in these NSI models in laboratory measurements, and especially extend the sensitivity reach in the searches of a dark photon to extremely low mass.

\acknowledgments
We would like to thank Joachim Kopp and Pedro A. N. Machado for useful discussions. This work was supported by the National Key Research and Development Program of China (Grants No. 2017YFA0402200 and No. 2022YFA1605000) and the National Natural Science Foundation of China (Grants No. 12175112, No. 12005111, and No. 11725522).
 
\bibliography{CDEX10Neutrino.bib}

%merlin.mbs apsrev4-1.bst 2010-07-25 4.21a (PWD, AO, DPC) hacked
%Control: key (0)
%Control: author (8) initials jnrlst
%Control: editor formatted (1) identically to author
%Control: production of article title (-1) disabled
%Control: page (0) single
%Control: year (1) truncated
%Control: production of eprint (0) enabled
\begin{thebibliography}{81}%
\makeatletter
\providecommand \@ifxundefined [1]{%
 \@ifx{#1\undefined}
}%
\providecommand \@ifnum [1]{%
 \ifnum #1\expandafter \@firstoftwo
 \else \expandafter \@secondoftwo
 \fi
}%
\providecommand \@ifx [1]{%
 \ifx #1\expandafter \@firstoftwo
 \else \expandafter \@secondoftwo
 \fi
}%
\providecommand \natexlab [1]{#1}%
\providecommand \enquote  [1]{``#1''}%
\providecommand \bibnamefont  [1]{#1}%
\providecommand \bibfnamefont [1]{#1}%
\providecommand \citenamefont [1]{#1}%
\providecommand \href@noop [0]{\@secondoftwo}%
\providecommand \href [0]{\begingroup \@sanitize@url \@href}%
\providecommand \@href[1]{\@@startlink{#1}\@@href}%
\providecommand \@@href[1]{\endgroup#1\@@endlink}%
\providecommand \@sanitize@url [0]{\catcode `\\12\catcode `\$12\catcode
  `\&12\catcode `\#12\catcode `\^12\catcode `\_12\catcode `\%12\relax}%
\providecommand \@@startlink[1]{}%
\providecommand \@@endlink[0]{}%
\providecommand \url  [0]{\begingroup\@sanitize@url \@url }%
\providecommand \@url [1]{\endgroup\@href {#1}{\urlprefix }}%
\providecommand \urlprefix  [0]{URL }%
\providecommand \Eprint [0]{\href }%
\providecommand \doibase [0]{http://dx.doi.org/}%
\providecommand \selectlanguage [0]{\@gobble}%
\providecommand \bibinfo  [0]{\@secondoftwo}%
\providecommand \bibfield  [0]{\@secondoftwo}%
\providecommand \translation [1]{[#1]}%
\providecommand \BibitemOpen [0]{}%
\providecommand \bibitemStop [0]{}%
\providecommand \bibitemNoStop [0]{.\EOS\space}%
\providecommand \EOS [0]{\spacefactor3000\relax}%
\providecommand \BibitemShut  [1]{\csname bibitem#1\endcsname}%
\let\auto@bib@innerbib\@empty
%</preamble>
\bibitem [{\citenamefont {Young}(2017)}]{cosmology}%
  \BibitemOpen
  \bibfield  {author} {\bibinfo {author} {\bibfnamefont {B.~L.}\ \bibnamefont
  {Young}},\ }\href {\doibase 10.1007/s11467-016-0583-4} {\bibfield  {journal}
  {\bibinfo  {journal} {Front. Phys.}\ }\textbf {\bibinfo {volume} {12}},\
  \bibinfo {eid} {121201} (\bibinfo {year} {2017})}\BibitemShut {NoStop}%
\bibitem [{\citenamefont {{Bhupal Dev}}\ \emph {et~al.}(2019)\citenamefont
  {{Bhupal Dev}} \emph {et~al.}}]{Proceedings:2019qno}%
  \BibitemOpen
  \bibfield  {author} {\bibinfo {author} {\bibfnamefont {P.~S.}\ \bibnamefont
  {{Bhupal Dev}}} \emph {et~al.},\ }\href {\doibase
  10.21468/SciPostPhysProc.2.001} {\bibfield  {journal} {\bibinfo  {journal}
  {SciPost Phys. Proc.}\ }\textbf {\bibinfo {volume} {2}},\ \bibinfo {pages}
  {001} (\bibinfo {year} {2019})}\BibitemShut {NoStop}%
\bibitem [{\citenamefont {Ohlsson}(2013)}]{Ohlsson_2013}%
  \BibitemOpen
  \bibfield  {author} {\bibinfo {author} {\bibfnamefont {T.}~\bibnamefont
  {Ohlsson}},\ }\href {\doibase 10.1088/0034-4885/76/4/044201} {\bibfield
  {journal} {\bibinfo  {journal} {Rep. Prog. Phys.}\ }\textbf {\bibinfo
  {volume} {76}},\ \bibinfo {pages} {044201} (\bibinfo {year}
  {2013})}\BibitemShut {NoStop}%
\bibitem [{\citenamefont {Ko}\ \emph {et~al.}(2017)\citenamefont {Ko} \emph
  {et~al.}}]{NEOS:PRL2017}%
  \BibitemOpen
  \bibfield  {author} {\bibinfo {author} {\bibfnamefont {Y.~J.}\ \bibnamefont
  {Ko}} \emph {et~al.} (\bibinfo {collaboration} {NEOS Collaboration}),\ }\href
  {\doibase 10.1103/PhysRevLett.118.121802} {\bibfield  {journal} {\bibinfo
  {journal} {Phys. Rev. Lett.}\ }\textbf {\bibinfo {volume} {118}},\ \bibinfo
  {pages} {121802} (\bibinfo {year} {2017})}\BibitemShut {NoStop}%
\bibitem [{\citenamefont {Serebrov}\ \emph {et~al.}(2021)\citenamefont
  {Serebrov} \emph {et~al.}}]{Neutrino4:PRD2021}%
  \BibitemOpen
  \bibfield  {author} {\bibinfo {author} {\bibfnamefont {A.~P.}\ \bibnamefont
  {Serebrov}} \emph {et~al.},\ }\href {\doibase 10.1103/PhysRevD.104.032003}
  {\bibfield  {journal} {\bibinfo  {journal} {Phys. Rev. D}\ }\textbf {\bibinfo
  {volume} {104}},\ \bibinfo {pages} {032003} (\bibinfo {year}
  {2021})}\BibitemShut {NoStop}%
\bibitem [{\citenamefont {Almaz\'an}\ \emph {et~al.}(2020)\citenamefont
  {Almaz\'an} \emph {et~al.}}]{STEREO:PRD2020}%
  \BibitemOpen
  \bibfield  {author} {\bibinfo {author} {\bibfnamefont {H.}~\bibnamefont
  {Almaz\'an}} \emph {et~al.} (\bibinfo {collaboration} {STEREO
  Collaboration}),\ }\href {\doibase 10.1103/PhysRevD.102.052002} {\bibfield
  {journal} {\bibinfo  {journal} {Phys. Rev. D}\ }\textbf {\bibinfo {volume}
  {102}},\ \bibinfo {pages} {052002} (\bibinfo {year} {2020})}\BibitemShut
  {NoStop}%
\bibitem [{\citenamefont {Ashenfelter}\ \emph {et~al.}(2018)\citenamefont
  {Ashenfelter} \emph {et~al.}}]{PROSPECT:PRL2018}%
  \BibitemOpen
  \bibfield  {author} {\bibinfo {author} {\bibfnamefont {J.}~\bibnamefont
  {Ashenfelter}} \emph {et~al.} (\bibinfo {collaboration} {PROSPECT
  Collaboration}),\ }\href {\doibase 10.1103/PhysRevLett.121.251802} {\bibfield
   {journal} {\bibinfo  {journal} {Phys. Rev. Lett.}\ }\textbf {\bibinfo
  {volume} {121}},\ \bibinfo {pages} {251802} (\bibinfo {year}
  {2018})}\BibitemShut {NoStop}%
\bibitem [{\citenamefont {Alekseev}\ \emph {et~al.}(2018)\citenamefont
  {Alekseev} \emph {et~al.}}]{DANSS:PLB2018}%
  \BibitemOpen
  \bibfield  {author} {\bibinfo {author} {\bibfnamefont {I.}~\bibnamefont
  {Alekseev}} \emph {et~al.},\ }\href {\doibase
  https://doi.org/10.1016/j.physletb.2018.10.038} {\bibfield  {journal}
  {\bibinfo  {journal} {Phys. Lett. B}\ }\textbf {\bibinfo {volume} {787}},\
  \bibinfo {pages} {56} (\bibinfo {year} {2018})}\BibitemShut {NoStop}%
\bibitem [{\citenamefont {Abreu}\ \emph {et~al.}(2021)\citenamefont {Abreu}
  \emph {et~al.}}]{Solid:JINST2021}%
  \BibitemOpen
  \bibfield  {author} {\bibinfo {author} {\bibfnamefont {Y.}~\bibnamefont
  {Abreu}} \emph {et~al.},\ }\href {\doibase 10.1088/1748-0221/16/02/p02025}
  {\bibfield  {journal} {\bibinfo  {journal} {J. Instrum.}\ }\textbf {\bibinfo
  {volume} {16}},\ \bibinfo {pages} {P02025} (\bibinfo {year}
  {2021})}\BibitemShut {NoStop}%
\bibitem [{\citenamefont {An}\ \emph {et~al.}(2014)\citenamefont {An} \emph
  {et~al.}}]{DayaBay:PRL2014}%
  \BibitemOpen
  \bibfield  {author} {\bibinfo {author} {\bibfnamefont {F.~P.}\ \bibnamefont
  {An}} \emph {et~al.} (\bibinfo {collaboration} {Daya Bay Collaboration}),\
  }\href {\doibase 10.1103/PhysRevLett.113.141802} {\bibfield  {journal}
  {\bibinfo  {journal} {Phys. Rev. Lett.}\ }\textbf {\bibinfo {volume} {113}},\
  \bibinfo {pages} {141802} (\bibinfo {year} {2014})}\BibitemShut {NoStop}%
\bibitem [{\citenamefont {Choi}\ \emph {et~al.}(2020)\citenamefont {Choi} \emph
  {et~al.}}]{RENO:PRL2020}%
  \BibitemOpen
  \bibfield  {author} {\bibinfo {author} {\bibfnamefont {J.~H.}\ \bibnamefont
  {Choi}} \emph {et~al.} (\bibinfo {collaboration} {RENO Collaboration}),\
  }\href {\doibase 10.1103/PhysRevLett.125.191801} {\bibfield  {journal}
  {\bibinfo  {journal} {Phys. Rev. Lett.}\ }\textbf {\bibinfo {volume} {125}},\
  \bibinfo {pages} {191801} (\bibinfo {year} {2020})}\BibitemShut {NoStop}%
\bibitem [{\citenamefont {Abusleme}\ \emph {et~al.}(2020)\citenamefont
  {Abusleme} \emph {et~al.}}]{JUNO:2020ijm}%
  \BibitemOpen
  \bibfield  {author} {\bibinfo {author} {\bibfnamefont {A.}~\bibnamefont
  {Abusleme}} \emph {et~al.} (\bibinfo {collaboration} {JUNO Collaboration}),\
  }\href@noop {} {\  (\bibinfo {year} {2020})},\ \Eprint
  {http://arxiv.org/abs/2005.08745} {arXiv:2005.08745 [physics.ins-det]}
  \BibitemShut {NoStop}%
\bibitem [{\citenamefont {Aguilar-Arevalo}\ \emph {et~al.}(2021)\citenamefont
  {Aguilar-Arevalo} \emph {et~al.}}]{MiniBooNE:PRD2021}%
  \BibitemOpen
  \bibfield  {author} {\bibinfo {author} {\bibfnamefont {A.~A.}\ \bibnamefont
  {Aguilar-Arevalo}} \emph {et~al.} (\bibinfo {collaboration} {MiniBooNE
  Collaboration}),\ }\href {\doibase 10.1103/PhysRevD.103.052002} {\bibfield
  {journal} {\bibinfo  {journal} {Phys. Rev. D}\ }\textbf {\bibinfo {volume}
  {103}},\ \bibinfo {pages} {052002} (\bibinfo {year} {2021})}\BibitemShut
  {NoStop}%
\bibitem [{\citenamefont {Aguilar}\ \emph {et~al.}(2001)\citenamefont {Aguilar}
  \emph {et~al.}}]{LSND:PRD2001}%
  \BibitemOpen
  \bibfield  {author} {\bibinfo {author} {\bibfnamefont {A.}~\bibnamefont
  {Aguilar}} \emph {et~al.} (\bibinfo {collaboration} {LSND Collaboration}),\
  }\href {\doibase 10.1103/PhysRevD.64.112007} {\bibfield  {journal} {\bibinfo
  {journal} {Phys. Rev. D}\ }\textbf {\bibinfo {volume} {64}},\ \bibinfo
  {pages} {112007} (\bibinfo {year} {2001})}\BibitemShut {NoStop}%
\bibitem [{\citenamefont {Abratenko}\ \emph {et~al.}(2022)\citenamefont
  {Abratenko} \emph {et~al.}}]{MicroBooNE}%
  \BibitemOpen
  \bibfield  {author} {\bibinfo {author} {\bibfnamefont {P.}~\bibnamefont
  {Abratenko}} \emph {et~al.} (\bibinfo {collaboration} {MicroBooNE
  Collaboration}),\ }\href {\doibase 10.1103/PhysRevLett.128.241801} {\bibfield
   {journal} {\bibinfo  {journal} {Phys. Rev. Lett.}\ }\textbf {\bibinfo
  {volume} {128}},\ \bibinfo {pages} {241801} (\bibinfo {year}
  {2022})}\BibitemShut {NoStop}%
\bibitem [{\citenamefont {Ajimura}\ \emph {et~al.}(2021)\citenamefont {Ajimura}
  \emph {et~al.}}]{JSNS2:NIMA2021}%
  \BibitemOpen
  \bibfield  {author} {\bibinfo {author} {\bibfnamefont {S.}~\bibnamefont
  {Ajimura}} \emph {et~al.},\ }\href {\doibase
  https://doi.org/10.1016/j.nima.2021.165742} {\bibfield  {journal} {\bibinfo
  {journal} {Nucl. Instrum. Methods Phys. Res., Sect. A}\ }\textbf {\bibinfo
  {volume} {1014}},\ \bibinfo {pages} {165742} (\bibinfo {year}
  {2021})}\BibitemShut {NoStop}%
\bibitem [{\citenamefont {Hampel}\ \emph {et~al.}(1998)\citenamefont {Hampel}
  \emph {et~al.}}]{GALLEX:PLB1998}%
  \BibitemOpen
  \bibfield  {author} {\bibinfo {author} {\bibfnamefont {W.}~\bibnamefont
  {Hampel}} \emph {et~al.},\ }\href {\doibase
  https://doi.org/10.1016/S0370-2693(97)01562-1} {\bibfield  {journal}
  {\bibinfo  {journal} {Phys. Lett. B}\ }\textbf {\bibinfo {volume} {420}},\
  \bibinfo {pages} {114} (\bibinfo {year} {1998})}\BibitemShut {NoStop}%
\bibitem [{\citenamefont {Abdurashitov}\ \emph {et~al.}(1999)\citenamefont
  {Abdurashitov} \emph {et~al.}}]{SAGE:PRC1999}%
  \BibitemOpen
  \bibfield  {author} {\bibinfo {author} {\bibfnamefont {J.~N.}\ \bibnamefont
  {Abdurashitov}} \emph {et~al.} (\bibinfo {collaboration} {SAGE
  Collaboration}),\ }\href {\doibase 10.1103/PhysRevC.59.2246} {\bibfield
  {journal} {\bibinfo  {journal} {Phys. Rev. C}\ }\textbf {\bibinfo {volume}
  {59}},\ \bibinfo {pages} {2246} (\bibinfo {year} {1999})}\BibitemShut
  {NoStop}%
\bibitem [{\citenamefont {Barinov}\ \emph
  {et~al.}(2022{\natexlab{a}})\citenamefont {Barinov} \emph
  {et~al.}}]{BEST:PRL2022}%
  \BibitemOpen
  \bibfield  {author} {\bibinfo {author} {\bibfnamefont {V.~V.}\ \bibnamefont
  {Barinov}} \emph {et~al.},\ }\href {\doibase 10.1103/PhysRevLett.128.232501}
  {\bibfield  {journal} {\bibinfo  {journal} {Phys. Rev. Lett.}\ }\textbf
  {\bibinfo {volume} {128}},\ \bibinfo {pages} {232501} (\bibinfo {year}
  {2022}{\natexlab{a}})}\BibitemShut {NoStop}%
\bibitem [{\citenamefont {Barinov}\ \emph
  {et~al.}(2022{\natexlab{b}})\citenamefont {Barinov} \emph {et~al.}}]{BEST1}%
  \BibitemOpen
  \bibfield  {author} {\bibinfo {author} {\bibfnamefont {V.~V.}\ \bibnamefont
  {Barinov}} \emph {et~al.},\ }\href {\doibase 10.1103/PhysRevC.105.065502}
  {\bibfield  {journal} {\bibinfo  {journal} {Phys. Rev. C}\ }\textbf {\bibinfo
  {volume} {105}},\ \bibinfo {pages} {065502} (\bibinfo {year}
  {2022}{\natexlab{b}})}\BibitemShut {NoStop}%
\bibitem [{\citenamefont {Barinov}\ and\ \citenamefont
  {Gorbunov}(2022)}]{BEST2}%
  \BibitemOpen
  \bibfield  {author} {\bibinfo {author} {\bibfnamefont {V.}~\bibnamefont
  {Barinov}}\ and\ \bibinfo {author} {\bibfnamefont {D.}~\bibnamefont
  {Gorbunov}},\ }\href {\doibase 10.1103/PhysRevD.105.L051703} {\bibfield
  {journal} {\bibinfo  {journal} {Phys. Rev. D}\ }\textbf {\bibinfo {volume}
  {105}},\ \bibinfo {pages} {L051703} (\bibinfo {year} {2022})}\BibitemShut
  {NoStop}%
\bibitem [{\citenamefont {Haxton}\ \emph {et~al.}(2013)\citenamefont {Haxton},
  \citenamefont {Hamish~Robertson},\ and\ \citenamefont
  {Serenelli}}]{solarneutrino}%
  \BibitemOpen
  \bibfield  {author} {\bibinfo {author} {\bibfnamefont {W.}~\bibnamefont
  {Haxton}}, \bibinfo {author} {\bibfnamefont {R.}~\bibnamefont
  {Hamish~Robertson}}, \ and\ \bibinfo {author} {\bibfnamefont {A.~M.}\
  \bibnamefont {Serenelli}},\ }\href {\doibase
  10.1146/annurev-astro-081811-125539} {\bibfield  {journal} {\bibinfo
  {journal} {Annu. Rev. Astron. Astrophys.}\ }\textbf {\bibinfo {volume}
  {51}},\ \bibinfo {pages} {21} (\bibinfo {year} {2013})}\BibitemShut {NoStop}%
\bibitem [{\citenamefont {Beacom}(2010)}]{SupernovaNeutrino}%
  \BibitemOpen
  \bibfield  {author} {\bibinfo {author} {\bibfnamefont {J.~F.}\ \bibnamefont
  {Beacom}},\ }\href {\doibase 10.1146/annurev.nucl.010909.083331} {\bibfield
  {journal} {\bibinfo  {journal} {Annu. Rev. Nucl. Part. Sci.}\ }\textbf
  {\bibinfo {volume} {60}},\ \bibinfo {pages} {439} (\bibinfo {year}
  {2010})}\BibitemShut {NoStop}%
\bibitem [{\citenamefont {Lunardini}(2016)}]{SupernovaNeutrino2}%
  \BibitemOpen
  \bibfield  {author} {\bibinfo {author} {\bibfnamefont {C.}~\bibnamefont
  {Lunardini}},\ }\href {\doibase
  https://doi.org/10.1016/j.astropartphys.2016.02.005} {\bibfield  {journal}
  {\bibinfo  {journal} {Astropart. Phys.}\ }\textbf {\bibinfo {volume} {79}},\
  \bibinfo {pages} {49} (\bibinfo {year} {2016})}\BibitemShut {NoStop}%
\bibitem [{\citenamefont {Kajita}(2012)}]{Kajita:2012vc}%
  \BibitemOpen
  \bibfield  {author} {\bibinfo {author} {\bibfnamefont {T.}~\bibnamefont
  {Kajita}},\ }\href {\doibase 10.1155/2012/504715} {\bibfield  {journal}
  {\bibinfo  {journal} {Adv. High Energy Phys.}\ }\textbf {\bibinfo {volume}
  {2012}},\ \bibinfo {pages} {504715} (\bibinfo {year} {2012})}\BibitemShut
  {NoStop}%
\bibitem [{\citenamefont {Formaggio}\ and\ \citenamefont
  {Zeller}(2012)}]{RevModPhys84}%
  \BibitemOpen
  \bibfield  {author} {\bibinfo {author} {\bibfnamefont {J.~A.}\ \bibnamefont
  {Formaggio}}\ and\ \bibinfo {author} {\bibfnamefont {G.~P.}\ \bibnamefont
  {Zeller}},\ }\href {\doibase 10.1103/RevModPhys.84.1307} {\bibfield
  {journal} {\bibinfo  {journal} {Rev. Mod. Phys.}\ }\textbf {\bibinfo {volume}
  {84}},\ \bibinfo {pages} {1307} (\bibinfo {year} {2012})}\BibitemShut
  {NoStop}%
\bibitem [{\citenamefont {Jiang}\ \emph {et~al.}(2018)\citenamefont {Jiang}
  \emph {et~al.}}]{cdex102018}%
  \BibitemOpen
  \bibfield  {author} {\bibinfo {author} {\bibfnamefont {H.}~\bibnamefont
  {Jiang}} \emph {et~al.} (\bibinfo {collaboration} {CDEX Collaboration}),\
  }\href {\doibase 10.1103/PhysRevLett.120.241301} {\bibfield  {journal}
  {\bibinfo  {journal} {Phys. Rev. Lett.}\ }\textbf {\bibinfo {volume} {120}},\
  \bibinfo {pages} {241301} (\bibinfo {year} {2018})}\BibitemShut {NoStop}%
\bibitem [{\citenamefont {She}\ \emph {et~al.}(2020)\citenamefont {She} \emph
  {et~al.}}]{cdex_darkphoton}%
  \BibitemOpen
  \bibfield  {author} {\bibinfo {author} {\bibfnamefont {Z.}~\bibnamefont
  {She}} \emph {et~al.} (\bibinfo {collaboration} {CDEX Collaboration}),\
  }\href {\doibase 10.1103/PhysRevLett.124.111301} {\bibfield  {journal}
  {\bibinfo  {journal} {Phys. Rev. Lett.}\ }\textbf {\bibinfo {volume} {124}},\
  \bibinfo {pages} {111301} (\bibinfo {year} {2020})}\BibitemShut {NoStop}%
\bibitem [{\citenamefont {Wang}\ \emph {et~al.}(2021)\citenamefont {Wang} \emph
  {et~al.}}]{cdex10_eft}%
  \BibitemOpen
  \bibfield  {author} {\bibinfo {author} {\bibfnamefont {Y.}~\bibnamefont
  {Wang}} \emph {et~al.} (\bibinfo {collaboration} {CDEX Collaboration}),\
  }\href {\doibase 10.1007/s11433-020-1666-8} {\bibfield  {journal} {\bibinfo
  {journal} {Sci. China Phys. Mech. Astron.}\ }\textbf {\bibinfo {volume}
  {64}},\ \bibinfo {pages} {281011} (\bibinfo {year} {2021})}\BibitemShut
  {NoStop}%
\bibitem [{\citenamefont {Xu}\ \emph {et~al.}(2022)\citenamefont {Xu} \emph
  {et~al.}}]{CRDM}%
  \BibitemOpen
  \bibfield  {author} {\bibinfo {author} {\bibfnamefont {R.}~\bibnamefont {Xu}}
  \emph {et~al.} (\bibinfo {collaboration} {CDEX Collaboration}),\ }\href
  {\doibase 10.1103/PhysRevD.106.052008} {\bibfield  {journal} {\bibinfo
  {journal} {Phys. Rev. D}\ }\textbf {\bibinfo {volume} {106}},\ \bibinfo
  {pages} {052008} (\bibinfo {year} {2022})}\BibitemShut {NoStop}%
\bibitem [{\citenamefont {Zhang}\ \emph {et~al.}(2022)\citenamefont {Zhang}
  \emph {et~al.}}]{CDEX_DM_e}%
  \BibitemOpen
  \bibfield  {author} {\bibinfo {author} {\bibfnamefont {Z.~Y.}\ \bibnamefont
  {Zhang}} \emph {et~al.} (\bibinfo {collaboration} {CDEX Collaboration}),\
  }\href {\doibase 10.1103/PhysRevLett.129.221301} {\bibfield  {journal}
  {\bibinfo  {journal} {Phys. Rev. Lett.}\ }\textbf {\bibinfo {volume} {129}},\
  \bibinfo {pages} {221301} (\bibinfo {year} {2022})}\BibitemShut {NoStop}%
\bibitem [{\citenamefont {Jiang}\ \emph {et~al.}(2019)\citenamefont {Jiang}
  \emph {et~al.}}]{cdex10_tech}%
  \BibitemOpen
  \bibfield  {author} {\bibinfo {author} {\bibfnamefont {H.}~\bibnamefont
  {Jiang}} \emph {et~al.} (\bibinfo {collaboration} {CDEX Collaboration}),\
  }\href {\doibase 10.1007/s11433-018-8001-3} {\bibfield  {journal} {\bibinfo
  {journal} {Sci. China Phys. Mech. Astron}\ }\textbf {\bibinfo {volume}
  {62}},\ \bibinfo {pages} {31012} (\bibinfo {year} {2019})}\BibitemShut
  {NoStop}%
\bibitem [{\citenamefont {Harnik}\ \emph {et~al.}(2012)\citenamefont {Harnik},
  \citenamefont {Kopp},\ and\ \citenamefont {Machado}}]{newphysics}%
  \BibitemOpen
  \bibfield  {author} {\bibinfo {author} {\bibfnamefont {R.}~\bibnamefont
  {Harnik}}, \bibinfo {author} {\bibfnamefont {J.}~\bibnamefont {Kopp}}, \ and\
  \bibinfo {author} {\bibfnamefont {P.~A.}\ \bibnamefont {Machado}},\ }\href
  {\doibase 10.1088/1475-7516/2012/07/026} {\bibfield  {journal} {\bibinfo
  {journal} {J. Cosmol. Astropart. Phys.}\ }\textbf {\bibinfo {volume} {07}},\
  \bibinfo {pages} {026} (\bibinfo {year} {2012})}\BibitemShut {NoStop}%
\bibitem [{\citenamefont {Bilmi\ifmmode~\mbox{\c{s}}\else \c{s}\fi{}}\ \emph
  {et~al.}(2015)\citenamefont {Bilmi\ifmmode~\mbox{\c{s}}\else \c{s}\fi{}}
  \emph {et~al.}}]{U1BL}%
  \BibitemOpen
  \bibfield  {author} {\bibinfo {author} {\bibfnamefont {S.}~\bibnamefont
  {Bilmi\ifmmode~\mbox{\c{s}}\else \c{s}\fi{}}} \emph {et~al.},\ }\href
  {\doibase 10.1103/PhysRevD.92.033009} {\bibfield  {journal} {\bibinfo
  {journal} {Phys. Rev. D}\ }\textbf {\bibinfo {volume} {92}},\ \bibinfo
  {pages} {033009} (\bibinfo {year} {2015})}\BibitemShut {NoStop}%
\bibitem [{\citenamefont {Fayet}(1980)}]{Fayet:1980ad}%
  \BibitemOpen
  \bibfield  {author} {\bibinfo {author} {\bibfnamefont {P.}~\bibnamefont
  {Fayet}},\ }\href {\doibase 10.1016/0370-2693(80)90488-8} {\bibfield
  {journal} {\bibinfo  {journal} {Phys. Lett. B}\ }\textbf {\bibinfo {volume}
  {95}},\ \bibinfo {pages} {285} (\bibinfo {year} {1980})}\BibitemShut
  {NoStop}%
\bibitem [{\citenamefont {Fayet}(1990)}]{Fayet:1990wx}%
  \BibitemOpen
  \bibfield  {author} {\bibinfo {author} {\bibfnamefont {P.}~\bibnamefont
  {Fayet}},\ }\href {\doibase 10.1016/0550-3213(90)90381-M} {\bibfield
  {journal} {\bibinfo  {journal} {Nucl. Phys. B}\ }\textbf {\bibinfo {volume}
  {347}},\ \bibinfo {pages} {743} (\bibinfo {year} {1990})}\BibitemShut
  {NoStop}%
\bibitem [{\citenamefont {Barbeau}\ \emph {et~al.}(2007)\citenamefont
  {Barbeau}, \citenamefont {Collar},\ and\ \citenamefont
  {Tench}}]{Barbeau_2007}%
  \BibitemOpen
  \bibfield  {author} {\bibinfo {author} {\bibfnamefont {P.~S.}\ \bibnamefont
  {Barbeau}}, \bibinfo {author} {\bibfnamefont {J.~I.}\ \bibnamefont {Collar}},
  \ and\ \bibinfo {author} {\bibfnamefont {O.}~\bibnamefont {Tench}},\ }\href
  {\doibase 10.1088/1475-7516/2007/09/009} {\bibfield  {journal} {\bibinfo
  {journal} {J. Cosmol. Astropart. Phys.}\ }\textbf {\bibinfo {volume} {09}},\
  \bibinfo {pages} {009} (\bibinfo {year} {2007})}\BibitemShut {NoStop}%
\bibitem [{\citenamefont {Soma}\ \emph {et~al.}(2016)\citenamefont {Soma} \emph
  {et~al.}}]{soma2016}%
  \BibitemOpen
  \bibfield  {author} {\bibinfo {author} {\bibfnamefont {A.}~\bibnamefont
  {Soma}} \emph {et~al.},\ }\href {\doibase 10.1016/j.nima.2016.08.044}
  {\bibfield  {journal} {\bibinfo  {journal} {Nucl. Instrum. Methods Phys.
  Res., Sect. A}\ }\textbf {\bibinfo {volume} {836}},\ \bibinfo {pages} {67 }
  (\bibinfo {year} {2016})}\BibitemShut {NoStop}%
\bibitem [{\citenamefont {Liu}\ \emph {et~al.}(2014)\citenamefont {Liu} \emph
  {et~al.}}]{cdex0}%
  \BibitemOpen
  \bibfield  {author} {\bibinfo {author} {\bibfnamefont {S.~K.}\ \bibnamefont
  {Liu}} \emph {et~al.} (\bibinfo {collaboration} {CDEX Collaboration}),\
  }\href {\doibase 10.1103/PhysRevD.90.032003} {\bibfield  {journal} {\bibinfo
  {journal} {Phys. Rev. D}\ }\textbf {\bibinfo {volume} {90}},\ \bibinfo
  {pages} {032003} (\bibinfo {year} {2014})}\BibitemShut {NoStop}%
\bibitem [{\citenamefont {Zhao}\ \emph {et~al.}(2013)\citenamefont {Zhao} \emph
  {et~al.}}]{cdex1}%
  \BibitemOpen
  \bibfield  {author} {\bibinfo {author} {\bibfnamefont {W.}~\bibnamefont
  {Zhao}} \emph {et~al.} (\bibinfo {collaboration} {CDEX Collaboration}),\
  }\href {\doibase 10.1103/PhysRevD.88.052004} {\bibfield  {journal} {\bibinfo
  {journal} {Phys. Rev. D}\ }\textbf {\bibinfo {volume} {88}},\ \bibinfo
  {pages} {052004} (\bibinfo {year} {2013})}\BibitemShut {NoStop}%
\bibitem [{\citenamefont {Yue}\ \emph {et~al.}(2014)\citenamefont {Yue} \emph
  {et~al.}}]{cdex12014}%
  \BibitemOpen
  \bibfield  {author} {\bibinfo {author} {\bibfnamefont {Q.}~\bibnamefont
  {Yue}} \emph {et~al.} (\bibinfo {collaboration} {CDEX Collaboration}),\
  }\href {\doibase 10.1103/PhysRevD.90.091701} {\bibfield  {journal} {\bibinfo
  {journal} {Phys. Rev. D}\ }\textbf {\bibinfo {volume} {90}},\ \bibinfo
  {pages} {091701} (\bibinfo {year} {2014})}\BibitemShut {NoStop}%
\bibitem [{\citenamefont {Zhao}\ \emph {et~al.}(2016)\citenamefont {Zhao} \emph
  {et~al.}}]{cdex12016}%
  \BibitemOpen
  \bibfield  {author} {\bibinfo {author} {\bibfnamefont {W.}~\bibnamefont
  {Zhao}} \emph {et~al.} (\bibinfo {collaboration} {CDEX Collaboration}),\
  }\href {\doibase 10.1103/PhysRevD.93.092003} {\bibfield  {journal} {\bibinfo
  {journal} {Phys. Rev. D}\ }\textbf {\bibinfo {volume} {93}},\ \bibinfo
  {pages} {092003} (\bibinfo {year} {2016})}\BibitemShut {NoStop}%
\bibitem [{\citenamefont {Yang}\ \emph
  {et~al.}(2018{\natexlab{a}})\citenamefont {Yang} \emph
  {et~al.}}]{cdex1b2018}%
  \BibitemOpen
  \bibfield  {author} {\bibinfo {author} {\bibfnamefont {L.~T.}\ \bibnamefont
  {Yang}} \emph {et~al.} (\bibinfo {collaboration} {CDEX Collaboration}),\
  }\href {\doibase 10.1088/1674-1137/42/2/023002} {\bibfield  {journal}
  {\bibinfo  {journal} {Chin. Phys. C}\ }\textbf {\bibinfo {volume} {42}},\
  \bibinfo {eid} {023002} (\bibinfo {year} {2018}{\natexlab{a}})}\BibitemShut
  {NoStop}%
\bibitem [{\citenamefont {Yang}\ \emph {et~al.}(2019)\citenamefont {Yang} \emph
  {et~al.}}]{cdex1b_am}%
  \BibitemOpen
  \bibfield  {author} {\bibinfo {author} {\bibfnamefont {L.~T.}\ \bibnamefont
  {Yang}} \emph {et~al.} (\bibinfo {collaboration} {CDEX Collaboration}),\
  }\href {\doibase 10.1103/PhysRevLett.123.221301} {\bibfield  {journal}
  {\bibinfo  {journal} {Phys. Rev. Lett.}\ }\textbf {\bibinfo {volume} {123}},\
  \bibinfo {pages} {221301} (\bibinfo {year} {2019})}\BibitemShut {NoStop}%
\bibitem [{\citenamefont {Cheng}\ \emph {et~al.}(2017)\citenamefont {Cheng}
  \emph {et~al.}}]{cjpl}%
  \BibitemOpen
  \bibfield  {author} {\bibinfo {author} {\bibfnamefont {J.~P.}\ \bibnamefont
  {Cheng}} \emph {et~al.},\ }\href {\doibase
  10.1146/annurev-nucl-102115-044842} {\bibfield  {journal} {\bibinfo
  {journal} {Annu. Rev. Nucl. Part. Sci.}\ }\textbf {\bibinfo {volume} {67}},\
  \bibinfo {pages} {231} (\bibinfo {year} {2017})}\BibitemShut {NoStop}%
\bibitem [{\citenamefont {Li}\ \emph {et~al.}(2014)\citenamefont {Li} \emph
  {et~al.}}]{Li:2014a}%
  \BibitemOpen
  \bibfield  {author} {\bibinfo {author} {\bibfnamefont {H.~B.}\ \bibnamefont
  {Li}} \emph {et~al.},\ }\href {\doibase
  https://doi.org/10.1016/j.astropartphys.2014.02.005} {\bibfield  {journal}
  {\bibinfo  {journal} {Astropart. Phys.}\ }\textbf {\bibinfo {volume} {56}},\
  \bibinfo {pages} {1} (\bibinfo {year} {2014})}\BibitemShut {NoStop}%
\bibitem [{\citenamefont {Yang}\ \emph
  {et~al.}(2018{\natexlab{b}})\citenamefont {Yang} \emph
  {et~al.}}]{Yang:2018a}%
  \BibitemOpen
  \bibfield  {author} {\bibinfo {author} {\bibfnamefont {L.~T.}\ \bibnamefont
  {Yang}} \emph {et~al.},\ }\href {\doibase
  https://doi.org/10.1016/j.nima.2017.12.078} {\bibfield  {journal} {\bibinfo
  {journal} {Nucl. Instrum. Methods Phys. Res., Sect. A}\ }\textbf {\bibinfo
  {volume} {886}},\ \bibinfo {pages} {13} (\bibinfo {year}
  {2018}{\natexlab{b}})}\BibitemShut {NoStop}%
\bibitem [{\citenamefont {Bahcall}(1963)}]{K-X-ray}%
  \BibitemOpen
  \bibfield  {author} {\bibinfo {author} {\bibfnamefont {J.~N.}\ \bibnamefont
  {Bahcall}},\ }\href {\doibase 10.1103/PhysRev.132.362} {\bibfield  {journal}
  {\bibinfo  {journal} {Phys. Rev.}\ }\textbf {\bibinfo {volume} {132}},\
  \bibinfo {pages} {362} (\bibinfo {year} {1963})}\BibitemShut {NoStop}%
\bibitem [{\citenamefont {Ziegler}\ \emph {et~al.}(2010)\citenamefont
  {Ziegler}, \citenamefont {Ziegler},\ and\ \citenamefont {Biersack}}]{TRIM}%
  \BibitemOpen
  \bibfield  {author} {\bibinfo {author} {\bibfnamefont {J.~F.}\ \bibnamefont
  {Ziegler}}, \bibinfo {author} {\bibfnamefont {M.}~\bibnamefont {Ziegler}}, \
  and\ \bibinfo {author} {\bibfnamefont {J.}~\bibnamefont {Biersack}},\ }\href
  {\doibase https://doi.org/10.1016/j.nimb.2010.02.091} {\bibfield  {journal}
  {\bibinfo  {journal} {Nucl. Instrum. Methods Phys. Res., Sect. B}\ }\textbf
  {\bibinfo {volume} {268}},\ \bibinfo {pages} {1818} (\bibinfo {year}
  {2010})}\BibitemShut {NoStop}%
\bibitem [{\citenamefont {Feldman}\ and\ \citenamefont
  {Cousins}(1998)}]{fcmethod}%
  \BibitemOpen
  \bibfield  {author} {\bibinfo {author} {\bibfnamefont {G.~J.}\ \bibnamefont
  {Feldman}}\ and\ \bibinfo {author} {\bibfnamefont {R.~D.}\ \bibnamefont
  {Cousins}},\ }\href {\doibase 10.1103/PhysRevD.57.3873} {\bibfield  {journal}
  {\bibinfo  {journal} {Phys. Rev. D}\ }\textbf {\bibinfo {volume} {57}},\
  \bibinfo {pages} {3873} (\bibinfo {year} {1998})}\BibitemShut {NoStop}%
\bibitem [{\citenamefont {Vinyoles}\ \emph {et~al.}(2017)\citenamefont
  {Vinyoles} \emph {et~al.}}]{Vinyoles2016}%
  \BibitemOpen
  \bibfield  {author} {\bibinfo {author} {\bibfnamefont {N.}~\bibnamefont
  {Vinyoles}} \emph {et~al.},\ }\href {\doibase 10.3847/1538-4357/835/2/202}
  {\bibfield  {journal} {\bibinfo  {journal} {Astrophys. J.}\ }\textbf
  {\bibinfo {volume} {835}},\ \bibinfo {pages} {202} (\bibinfo {year}
  {2017})}\BibitemShut {NoStop}%
\bibitem [{\citenamefont {Engel}(1991)}]{form}%
  \BibitemOpen
  \bibfield  {author} {\bibinfo {author} {\bibfnamefont {J.}~\bibnamefont
  {Engel}},\ }\href {\doibase 10.1016/0370-2693(91)90712-Y} {\bibfield
  {journal} {\bibinfo  {journal} {Phys. Lett. B}\ }\textbf {\bibinfo {volume}
  {264}},\ \bibinfo {pages} {114} (\bibinfo {year} {1991})}\BibitemShut
  {NoStop}%
\bibitem [{\citenamefont {Lewin}\ and\ \citenamefont
  {Smith}(1996)}]{formfactor}%
  \BibitemOpen
  \bibfield  {author} {\bibinfo {author} {\bibfnamefont {J.}~\bibnamefont
  {Lewin}}\ and\ \bibinfo {author} {\bibfnamefont {P.}~\bibnamefont {Smith}},\
  }\href {\doibase 10.1016/S0927-6505(96)00047-3} {\bibfield  {journal}
  {\bibinfo  {journal} {Astropart. Phys.}\ }\textbf {\bibinfo {volume} {6}},\
  \bibinfo {pages} {87 } (\bibinfo {year} {1996})}\BibitemShut {NoStop}%
\bibitem [{\citenamefont {Deniz}\ \emph {et~al.}(2010)\citenamefont {Deniz}
  \emph {et~al.}}]{TEXONO_CsI}%
  \BibitemOpen
  \bibfield  {author} {\bibinfo {author} {\bibfnamefont {M.}~\bibnamefont
  {Deniz}} \emph {et~al.} (\bibinfo {collaboration} {TEXONO Collaboration}),\
  }\href {\doibase 10.1103/PhysRevD.81.072001} {\bibfield  {journal} {\bibinfo
  {journal} {Phys. Rev. D}\ }\textbf {\bibinfo {volume} {81}},\ \bibinfo
  {pages} {072001} (\bibinfo {year} {2010})}\BibitemShut {NoStop}%
\bibitem [{\citenamefont {Bellini}\ \emph {et~al.}(2011)\citenamefont {Bellini}
  \emph {et~al.}}]{Borexino_exclusion}%
  \BibitemOpen
  \bibfield  {author} {\bibinfo {author} {\bibfnamefont {G.}~\bibnamefont
  {Bellini}} \emph {et~al.} (\bibinfo {collaboration} {Borexino
  Collaboration}),\ }\href {\doibase 10.1103/PhysRevLett.107.141302} {\bibfield
   {journal} {\bibinfo  {journal} {Phys. Rev. Lett.}\ }\textbf {\bibinfo
  {volume} {107}},\ \bibinfo {pages} {141302} (\bibinfo {year}
  {2011})}\BibitemShut {NoStop}%
\bibitem [{\citenamefont {Abe}\ \emph {et~al.}(2020)\citenamefont {Abe} \emph
  {et~al.}}]{XMASS}%
  \BibitemOpen
  \bibfield  {author} {\bibinfo {author} {\bibfnamefont {K.}~\bibnamefont
  {Abe}} \emph {et~al.} (\bibinfo {collaboration} {XMASS Collaboration}),\
  }\href {\doibase https://doi.org/10.1016/j.physletb.2020.135741} {\bibfield
  {journal} {\bibinfo  {journal} {Phys. Lett. B}\ }\textbf {\bibinfo {volume}
  {809}},\ \bibinfo {pages} {135741} (\bibinfo {year} {2020})}\BibitemShut
  {NoStop}%
\bibitem [{\citenamefont {Vilain}\ \emph {et~al.}(1993)\citenamefont {Vilain}
  \emph {et~al.}}]{VILAIN1993}%
  \BibitemOpen
  \bibfield  {author} {\bibinfo {author} {\bibfnamefont {P.}~\bibnamefont
  {Vilain}} \emph {et~al.} (\bibinfo {collaboration} {CHARM-II
  Collaboration}),\ }\href {\doibase
  https://doi.org/10.1016/0370-2693(93)90408-A} {\bibfield  {journal} {\bibinfo
   {journal} {Phys. Lett. B}\ }\textbf {\bibinfo {volume} {302}},\ \bibinfo
  {pages} {351} (\bibinfo {year} {1993})}\BibitemShut {NoStop}%
\bibitem [{\citenamefont {Vilain}\ \emph {et~al.}(1994)\citenamefont {Vilain}
  \emph {et~al.}}]{VILAIN1994}%
  \BibitemOpen
  \bibfield  {author} {\bibinfo {author} {\bibfnamefont {P.}~\bibnamefont
  {Vilain}} \emph {et~al.} (\bibinfo {collaboration} {CHARM-II
  Collaboration}),\ }\href {\doibase
  https://doi.org/10.1016/0370-2693(94)91421-4} {\bibfield  {journal} {\bibinfo
   {journal} {Phys. Lett. B}\ }\textbf {\bibinfo {volume} {335}},\ \bibinfo
  {pages} {246} (\bibinfo {year} {1994})}\BibitemShut {NoStop}%
\bibitem [{\citenamefont {Beda}\ \emph {et~al.}(2010)\citenamefont {Beda} \emph
  {et~al.}}]{GEMMA}%
  \BibitemOpen
  \bibfield  {author} {\bibinfo {author} {\bibfnamefont {A.~G.}\ \bibnamefont
  {Beda}} \emph {et~al.},\ }\href {\doibase 10.1134/S1547477110060063}
  {\bibfield  {journal} {\bibinfo  {journal} {Phys. Part. Nucl. Lett.}\
  }\textbf {\bibinfo {volume} {7}},\ \bibinfo {pages} {406} (\bibinfo {year}
  {2010})}\BibitemShut {NoStop}%
\bibitem [{\citenamefont {Andreev}\ \emph {et~al.}(2022)\citenamefont {Andreev}
  \emph {et~al.}}]{NA64e}%
  \BibitemOpen
  \bibfield  {author} {\bibinfo {author} {\bibfnamefont {Y.~M.}\ \bibnamefont
  {Andreev}} \emph {et~al.} (\bibinfo {collaboration} {NA64 Collaboration}),\
  }\href {\doibase 10.1103/PhysRevLett.129.161801} {\bibfield  {journal}
  {\bibinfo  {journal} {Phys. Rev. Lett.}\ }\textbf {\bibinfo {volume} {129}},\
  \bibinfo {pages} {161801} (\bibinfo {year} {2022})}\BibitemShut {NoStop}%
\bibitem [{\citenamefont {Redondo}(2008)}]{Sun1}%
  \BibitemOpen
  \bibfield  {author} {\bibinfo {author} {\bibfnamefont {J.}~\bibnamefont
  {Redondo}},\ }\href {\doibase 10.1088/1475-7516/2008/07/008} {\bibfield
  {journal} {\bibinfo  {journal} {J. Cosmol. Astropart. Phys.}\ }\textbf
  {\bibinfo {volume} {07}},\ \bibinfo {pages} {008} (\bibinfo {year}
  {2008})}\BibitemShut {NoStop}%
\bibitem [{\citenamefont {Raffelt}\ and\ \citenamefont
  {Starkman}(1989)}]{Sun2}%
  \BibitemOpen
  \bibfield  {author} {\bibinfo {author} {\bibfnamefont {G.~G.}\ \bibnamefont
  {Raffelt}}\ and\ \bibinfo {author} {\bibfnamefont {G.~D.}\ \bibnamefont
  {Starkman}},\ }\href {\doibase 10.1103/PhysRevD.40.942} {\bibfield  {journal}
  {\bibinfo  {journal} {Phys. Rev. D}\ }\textbf {\bibinfo {volume} {40}},\
  \bibinfo {pages} {942} (\bibinfo {year} {1989})}\BibitemShut {NoStop}%
\bibitem [{\citenamefont {Arik}\ \emph {et~al.}(2009)\citenamefont {Arik} \emph
  {et~al.}}]{CAST}%
  \BibitemOpen
  \bibfield  {author} {\bibinfo {author} {\bibfnamefont {E.}~\bibnamefont
  {Arik}} \emph {et~al.} (\bibinfo {collaboration} {CAST Collaboration}),\
  }\href {\doibase 10.1088/1475-7516/2009/02/008} {\bibfield  {journal}
  {\bibinfo  {journal} {J. Cosmol. Astropart. Phys.}\ }\textbf {\bibinfo
  {volume} {02}},\ \bibinfo {pages} {008} (\bibinfo {year} {2009})}\BibitemShut
  {NoStop}%
\bibitem [{\citenamefont {Mirizzi}\ \emph {et~al.}(2009)\citenamefont
  {Mirizzi}, \citenamefont {Redondo},\ and\ \citenamefont {Sigl}}]{CMB1}%
  \BibitemOpen
  \bibfield  {author} {\bibinfo {author} {\bibfnamefont {A.}~\bibnamefont
  {Mirizzi}}, \bibinfo {author} {\bibfnamefont {J.}~\bibnamefont {Redondo}}, \
  and\ \bibinfo {author} {\bibfnamefont {G.}~\bibnamefont {Sigl}},\ }\href
  {\doibase 10.1088/1475-7516/2009/03/026} {\bibfield  {journal} {\bibinfo
  {journal} {J. Cosmol. Astropart. Phys.}\ }\textbf {\bibinfo {volume} {03}},\
  \bibinfo {pages} {026} (\bibinfo {year} {2009})}\BibitemShut {NoStop}%
\bibitem [{\citenamefont {Fixsen}\ \emph {et~al.}(1996)\citenamefont {Fixsen}
  \emph {et~al.}}]{CMB2}%
  \BibitemOpen
  \bibfield  {author} {\bibinfo {author} {\bibfnamefont {D.~J.}\ \bibnamefont
  {Fixsen}} \emph {et~al.},\ }\href {\doibase 10.1086/178173} {\bibfield
  {journal} {\bibinfo  {journal} {Astrophys. J.}\ }\textbf {\bibinfo {volume}
  {473}},\ \bibinfo {pages} {576} (\bibinfo {year} {1996})}\BibitemShut
  {NoStop}%
\bibitem [{\citenamefont {Dent}\ \emph {et~al.}(2012)\citenamefont {Dent},
  \citenamefont {Ferrer},\ and\ \citenamefont {Krauss}}]{SN1987A}%
  \BibitemOpen
  \bibfield  {author} {\bibinfo {author} {\bibfnamefont {J.~B.}\ \bibnamefont
  {Dent}}, \bibinfo {author} {\bibfnamefont {F.}~\bibnamefont {Ferrer}}, \ and\
  \bibinfo {author} {\bibfnamefont {L.~M.}\ \bibnamefont {Krauss}},\
  }\href@noop {} {\  (\bibinfo {year} {2012})},\ \Eprint
  {http://arxiv.org/abs/1201.2683} {arXiv:1201.2683 [astro-ph.CO]} \BibitemShut
  {NoStop}%
\bibitem [{\citenamefont {Davidson}\ \emph {et~al.}(2000)\citenamefont
  {Davidson}, \citenamefont {Hannestad},\ and\ \citenamefont
  {Raffelt}}]{ENERGY_LOSS}%
  \BibitemOpen
  \bibfield  {author} {\bibinfo {author} {\bibfnamefont {S.}~\bibnamefont
  {Davidson}}, \bibinfo {author} {\bibfnamefont {S.}~\bibnamefont {Hannestad}},
  \ and\ \bibinfo {author} {\bibfnamefont {G.}~\bibnamefont {Raffelt}},\ }\href
  {\doibase 10.1088/1126-6708/2000/05/003} {\bibfield  {journal} {\bibinfo
  {journal} {J. High Energy Phys.}\ }\textbf {\bibinfo {volume} {05}},\
  \bibinfo {pages} {003} (\bibinfo {year} {2000})}\BibitemShut {NoStop}%
\bibitem [{\citenamefont {Ahlers}\ \emph {et~al.}(2008)\citenamefont {Ahlers}
  \emph {et~al.}}]{LSW}%
  \BibitemOpen
  \bibfield  {author} {\bibinfo {author} {\bibfnamefont {M.}~\bibnamefont
  {Ahlers}} \emph {et~al.},\ }\href {\doibase 10.1103/PhysRevD.77.095001}
  {\bibfield  {journal} {\bibinfo  {journal} {Phys. Rev. D}\ }\textbf {\bibinfo
  {volume} {77}},\ \bibinfo {pages} {095001} (\bibinfo {year}
  {2008})}\BibitemShut {NoStop}%
\bibitem [{\citenamefont {Bartlett}\ and\ \citenamefont
  {Loegl}(1988)}]{atomic_physics}%
  \BibitemOpen
  \bibfield  {author} {\bibinfo {author} {\bibfnamefont {D.~F.}\ \bibnamefont
  {Bartlett}}\ and\ \bibinfo {author} {\bibfnamefont {S.}~\bibnamefont
  {Loegl}},\ }\href {\doibase 10.1103/PhysRevLett.61.2285} {\bibfield
  {journal} {\bibinfo  {journal} {Phys. Rev. Lett.}\ }\textbf {\bibinfo
  {volume} {61}},\ \bibinfo {pages} {2285} (\bibinfo {year}
  {1988})}\BibitemShut {NoStop}%
\bibitem [{\citenamefont {Bordag}\ \emph {et~al.}(2001)\citenamefont {Bordag},
  \citenamefont {Mohideen},\ and\ \citenamefont {Mostepanenko}}]{fifth_force}%
  \BibitemOpen
  \bibfield  {author} {\bibinfo {author} {\bibfnamefont {M.}~\bibnamefont
  {Bordag}}, \bibinfo {author} {\bibfnamefont {U.}~\bibnamefont {Mohideen}}, \
  and\ \bibinfo {author} {\bibfnamefont {V.~M.}\ \bibnamefont {Mostepanenko}},\
  }\href {\doibase 10.1016/S0370-1573(01)00015-1} {\bibfield  {journal}
  {\bibinfo  {journal} {Phys. Rep.}\ }\textbf {\bibinfo {volume} {353}},\
  \bibinfo {pages} {1} (\bibinfo {year} {2001})}\BibitemShut {NoStop}%
\bibitem [{\citenamefont {Bjorken}\ \emph {et~al.}(2009)\citenamefont {Bjorken}
  \emph {et~al.}}]{fixed_1}%
  \BibitemOpen
  \bibfield  {author} {\bibinfo {author} {\bibfnamefont {J.~D.}\ \bibnamefont
  {Bjorken}} \emph {et~al.},\ }\href {\doibase 10.1103/PhysRevD.80.075018}
  {\bibfield  {journal} {\bibinfo  {journal} {Phys. Rev. D}\ }\textbf {\bibinfo
  {volume} {80}},\ \bibinfo {pages} {075018} (\bibinfo {year}
  {2009})}\BibitemShut {NoStop}%
\bibitem [{\citenamefont {Batell}\ \emph {et~al.}(2009)\citenamefont {Batell},
  \citenamefont {Pospelov},\ and\ \citenamefont {Ritz}}]{fixed_2}%
  \BibitemOpen
  \bibfield  {author} {\bibinfo {author} {\bibfnamefont {B.}~\bibnamefont
  {Batell}}, \bibinfo {author} {\bibfnamefont {M.}~\bibnamefont {Pospelov}}, \
  and\ \bibinfo {author} {\bibfnamefont {A.}~\bibnamefont {Ritz}},\ }\href
  {\doibase 10.1103/PhysRevD.80.095024} {\bibfield  {journal} {\bibinfo
  {journal} {Phys. Rev. D}\ }\textbf {\bibinfo {volume} {80}},\ \bibinfo
  {pages} {095024} (\bibinfo {year} {2009})}\BibitemShut {NoStop}%
\bibitem [{\citenamefont {Essig}\ \emph {et~al.}(2010)\citenamefont {Essig}
  \emph {et~al.}}]{fixed_3}%
  \BibitemOpen
  \bibfield  {author} {\bibinfo {author} {\bibfnamefont {R.}~\bibnamefont
  {Essig}} \emph {et~al.},\ }\href {\doibase 10.1103/PhysRevD.82.113008}
  {\bibfield  {journal} {\bibinfo  {journal} {Phys. Rev. D}\ }\textbf {\bibinfo
  {volume} {82}},\ \bibinfo {pages} {113008} (\bibinfo {year}
  {2010})}\BibitemShut {NoStop}%
\bibitem [{\citenamefont {Pospelov}(2009)}]{g-21}%
  \BibitemOpen
  \bibfield  {author} {\bibinfo {author} {\bibfnamefont {M.}~\bibnamefont
  {Pospelov}},\ }\href {\doibase 10.1103/PhysRevD.80.095002} {\bibfield
  {journal} {\bibinfo  {journal} {Phys. Rev. D}\ }\textbf {\bibinfo {volume}
  {80}},\ \bibinfo {pages} {095002} (\bibinfo {year} {2009})}\BibitemShut
  {NoStop}%
\bibitem [{\citenamefont {Bennett}\ \emph {et~al.}(2006)\citenamefont {Bennett}
  \emph {et~al.}}]{g-22}%
  \BibitemOpen
  \bibfield  {author} {\bibinfo {author} {\bibfnamefont {G.~W.}\ \bibnamefont
  {Bennett}} \emph {et~al.} (\bibinfo {collaboration} {Muon $g-2$
  Collaboration}),\ }\href {\doibase 10.1103/PhysRevD.73.072003} {\bibfield
  {journal} {\bibinfo  {journal} {Phys. Rev. D}\ }\textbf {\bibinfo {volume}
  {73}},\ \bibinfo {pages} {072003} (\bibinfo {year} {2006})}\BibitemShut
  {NoStop}%
\bibitem [{\citenamefont {Essig}\ \emph {et~al.}(2009)\citenamefont {Essig},
  \citenamefont {Schuster},\ and\ \citenamefont {Toro}}]{Upsilon}%
  \BibitemOpen
  \bibfield  {author} {\bibinfo {author} {\bibfnamefont {R.}~\bibnamefont
  {Essig}}, \bibinfo {author} {\bibfnamefont {P.}~\bibnamefont {Schuster}}, \
  and\ \bibinfo {author} {\bibfnamefont {N.}~\bibnamefont {Toro}},\ }\href
  {\doibase 10.1103/PhysRevD.80.015003} {\bibfield  {journal} {\bibinfo
  {journal} {Phys. Rev. D}\ }\textbf {\bibinfo {volume} {80}},\ \bibinfo
  {pages} {015003} (\bibinfo {year} {2009})}\BibitemShut {NoStop}%
\bibitem [{\citenamefont {Danilov}\ \emph {et~al.}(2019)\citenamefont {Danilov}
  \emph {et~al.}}]{TEXONO_DP}%
  \BibitemOpen
  \bibfield  {author} {\bibinfo {author} {\bibfnamefont {M.}~\bibnamefont
  {Danilov}} \emph {et~al.},\ }\href {\doibase 10.1103/PhysRevLett.122.041801}
  {\bibfield  {journal} {\bibinfo  {journal} {Phys. Rev. Lett.}\ }\textbf
  {\bibinfo {volume} {122}},\ \bibinfo {pages} {041801} (\bibinfo {year}
  {2019})}\BibitemShut {NoStop}%
\bibitem [{\citenamefont {Aguilar-Arevalo}\ \emph {et~al.}(2019)\citenamefont
  {Aguilar-Arevalo} \emph {et~al.}}]{DAMIC_DP}%
  \BibitemOpen
  \bibfield  {author} {\bibinfo {author} {\bibfnamefont {A.}~\bibnamefont
  {Aguilar-Arevalo}} \emph {et~al.} (\bibinfo {collaboration} {DAMIC
  Collaboration}),\ }\href {\doibase 10.1103/PhysRevLett.123.181802} {\bibfield
   {journal} {\bibinfo  {journal} {Phys. Rev. Lett.}\ }\textbf {\bibinfo
  {volume} {123}},\ \bibinfo {pages} {181802} (\bibinfo {year}
  {2019})}\BibitemShut {NoStop}%
\bibitem [{\citenamefont {Aralis}\ \emph {et~al.}(2020)\citenamefont {Aralis}
  \emph {et~al.}}]{SuperCDMS_DP}%
  \BibitemOpen
  \bibfield  {author} {\bibinfo {author} {\bibfnamefont {T.}~\bibnamefont
  {Aralis}} \emph {et~al.} (\bibinfo {collaboration} {SuperCDMS
  Collaboration}),\ }\href {\doibase 10.1103/PhysRevD.101.052008} {\bibfield
  {journal} {\bibinfo  {journal} {Phys. Rev. D}\ }\textbf {\bibinfo {volume}
  {101}},\ \bibinfo {pages} {052008} (\bibinfo {year} {2020})}\BibitemShut
  {NoStop}%
\bibitem [{\citenamefont {Aprile}\ \emph {et~al.}(2020)\citenamefont {Aprile}
  \emph {et~al.}}]{XENON1T_S1_S2_DP}%
  \BibitemOpen
  \bibfield  {author} {\bibinfo {author} {\bibfnamefont {E.}~\bibnamefont
  {Aprile}} \emph {et~al.} (\bibinfo {collaboration} {XENON Collaboration}),\
  }\href {\doibase 10.1103/PhysRevD.102.072004} {\bibfield  {journal} {\bibinfo
   {journal} {Phys. Rev. D}\ }\textbf {\bibinfo {volume} {102}},\ \bibinfo
  {pages} {072004} (\bibinfo {year} {2020})}\BibitemShut {NoStop}%
\bibitem [{\citenamefont {Aprile}\ \emph {et~al.}(2019)\citenamefont {Aprile}
  \emph {et~al.}}]{XENON1T_S2_DP}%
  \BibitemOpen
  \bibfield  {author} {\bibinfo {author} {\bibfnamefont {E.}~\bibnamefont
  {Aprile}} \emph {et~al.} (\bibinfo {collaboration} {XENON Collaboration}),\
  }\href {\doibase 10.1103/PhysRevLett.123.251801} {\bibfield  {journal}
  {\bibinfo  {journal} {Phys. Rev. Lett.}\ }\textbf {\bibinfo {volume} {123}},\
  \bibinfo {pages} {251801} (\bibinfo {year} {2019})}\BibitemShut {NoStop}%
\end{thebibliography}%

\end{document}